\documentclass[12pt]{article}
\usepackage{jheppub}

\usepackage{ amssymb,amsfonts,amsthm,amsmath}
\usepackage{epsfig}
\usepackage{graphicx}
\usepackage{subfigure}

\makeatletter

\@addtoreset{equation}{section}
\makeatother

\newcommand{\beq}{\begin{equation}}
\newcommand{\eeq}{\end{equation}}

\begin{document}



\begin{titlepage}

\setcounter{page}{0}

\renewcommand{\thefootnote}{\fnsymbol{footnote}}

\setcounter{page}{1} 


\begin{center}
{\LARGE \bf
Refined Chern-Simons Theory
\vskip 1cm
and
\vskip 1 cm
 Topological String}
\vskip 1.5cm

{\large
Mina Aganagic$^{1,2}$ and Shamil Shakirov$^{1,2,3}$
\\
\medskip
}

\vskip 0.5cm

{
\it
$^1$Center for Theoretical Physics, University of California, Berkeley, CA 94720, USA\\
\medskip
$^2$Department of Mathematics, University of California, Berkeley, CA 94720, USA\\
\medskip
$^3$Institute for Theoretical and Experimental Physics (ITEP), Moscow, Russia\\
\medskip
}
\end{center}

\centerline{{\bf Abstract}}
\medskip
\noindent

We show that refined Chern-Simons theory and large $N$ duality can be used to study the refined topological string with and without branes.
We derive the refined topological vertex of \cite{CIV} and \cite{AK} from a link invariant of the refined $SU(N)$ Chern-Simons theory on $S^3$, at infinite $N$. Quiver-like Chern-Simons theories, arising from Calabi-Yau manifolds with branes wrapped on several minimal $S^3$'s, give a dual description of a large class of toric Calabi-Yau.
We use this to derive the refined topological string amplitudes on a toric Calabi-Yau containing a shrinking ${\mathbb P}^2$ surface.  The result is suggestive of the refined topological vertex formalism for arbitrary toric Calabi-Yau manifolds in terms of a pair of vertices and a choice of a Morse flow on the toric graph, determining the vertex decomposition.  The dependence on the flow is reminiscent of the approach to the refined topological string in \cite{NO2}. As a byproduct, we show that large $N$ duality of the refined topological string explains the ``mirror symmetry`` of the refined colored HOMFLY invariants of knots.
\end{titlepage}
\setcounter{page}{1} 

\section{Introduction}

The topological string variant of the gauge/gravity duality \cite{GV} relates $SU(N)$ Chern-Simons theory on $S^3$ to the topological string on the conifold with ${\mathbb P}^1$ of size $Ng_s$. Chern-Simons gauge theory is the string field theory of topological D-branes on $S^3$ \cite{Witten:1992fb}.  The large $N$ duality in this case has an interpretation as a geometric transition that shrinks the ${S}^3$ and grows the ${\mathbb P}^1$. This was generalized in \cite{AMV, AKMV} to quiver-like Chern-Simons theories, dual to a larger class of Calabi-Yau manifolds. The equivalence can be used to obtain the {\it exact}, all genus closed string amplitudes on the dual Calabi-Yau. Eventually, Chern-Simons theory and large $N$ duality lead to topological vertex formalism \cite{TV} and a complete solution of topological string on toric Calabi-Yau manifolds.

As shown in \cite{GVI, GVII, OV}, topological A-model string can be interpreted as computing the index of M-theory on the Calabi-Yau, in the self-dual $\Omega$-background \cite{N, NO}. The M-theory partition function in the arbitrary $\Omega$-background defines the partition function of the refined topological string \cite{HIV, Dijkgraaf2006}.
D-branes of the topological string map to M5 branes wrapping Lagrangian submanifolds of the Calabi-Yau. The partition function of M5 branes in the general $\Omega$-background defines the partition function of the refined Chern-Simons theory \cite{AS}. The refined $SU(N)$ Chern-Simons theory was formulated and solved in \cite{AS, AS2}, and for arbitrary ADE gauge groups in \cite{Aganagic2012}.
Large $N$ dual of the refined $SU(N)$ Chern-Simons partition function on $S^3$ was shown in \cite{AS} to be the refined topological string partition function on the conifold. Moreover, the link invariants of the refined Chern-Simons theory are expected to compute refined topological string amplitudes on the conifold, together with non-compact branes.\footnote{Mathematically, the link invariants of Chern-Simons theory at large $N$ are conjectured in \cite{AS} to be related to the categorification of the HOMFLY polynomial. The conjecture was verified in many cases \cite{AS,O1, Superpoly2, Moscow1,  FGS, O2}. For earlier work on extracting the categorified HOMFLY from string theory, see \cite{GIV}.}
This opens up a path to generalize the approach to topological string of \cite{AMV, TV} to the refined case.

We first show how to derive the refined topological vertex of \cite{CIV, AK} from the refined Chern-Simons theory at infinite $N$\footnote{That the vertices of \cite{CIV} and \cite{AK} are equivalent, related by a change of basis, was shown recently in \cite{AwataFeigin}.}. The topological vertex is the topological string partition function on ${\mathbb C}^3$ with three stacks of Lagrangian D-branes.
At infinite $N$, the dual of geometry becomes, instead of the conifold, simply ${\mathbb C}^3$. The configuration of branes needed on ${\mathbb C}^3$ corresponds to a simple three-component link in $S^3$.
We will show that, following the same steps as in \cite{TV}, just in the refined context, we can derive the refined topological vertex of \cite{CIV} and \cite{AK}.  The key ingredient, which we derive, are the refined Ooguri-Vafa operators \cite{OV} that translate refined topological string amplitudes into knot invariants of refined Chern-Simons theory.

Next, we generalize the work of \cite{AMV} to the refined case.\footnote{Refined Chern-Simons theory also played an important role in solving the refined topological string on local Riemann surfaces, and in proving the refined OSV conjecture \cite{OSV,Aganagic:2012si}, in this case.} This corresponds to more general geometric transitions where more than one $S^3$ shrinks. The gauge theory side in this case is captured by a quiver.  The nodes correspond to $S^3$'s (or orbifolds of $S^3$'s) wrapped by M5 branes. The bi-fundamental matter comes from M2 branes wrapping holomorphic curves with boundaries on the three manifolds. The M-theory index is captured by computations of link invariants in the refined Chern-Simons theory, where links come from M2 brane boundaries, and the linking pattern is dictated by the Calabi-Yau geometry. The large $N$ duality implies that the partition function of these theories should be the same as the partition function of the refined topological string on the dual Calabi-Yau, obtained by shrinking the $S^3$'s and growing the ${\mathbb P}^1$'s in their place. As an illustration,\footnote{While this paper was in preparation, we were informed of the upcoming work by Amer Iqbal and Can Kozcaz which may overlap with our results on local ${\mathbb P}^2$ amplitudes. We are grateful to the authors for agreeing to coordinate submission with us.} we use this to solve the refined topological string on a Calabi-Yau containing the local ${\mathbb P}^2$.

To obtain refined topological string partition functions on an arbitrary toric Calabi-Yau manifold, we need to generalize the topological vertex formalism of \cite{TV}.  A formalism that applies to a restricted class of toric Calabi-Yau manifolds whose toric diagrams admit a preferred direction was proposed in \cite{AK, CIV}. However, the refined topological string exists on arbitrary toric Calabi-Yau manifolds.
The local ${\mathbb P}^2$ geometry which we studied using large $N$ duality falls outside of this class. We show that the refined partition function we derived admits a vertex decomposition, and one that is very suggestive of what the general refined topological vertex formalism may be. We will flesh out some of its ingredients. To define the refined topological string requires a choice of a $U(1)$ action on $X$ (in fact, this is a $U(1)_R$ symmetry in M-theory). Different choices of this can be thought of as specifying a Morse flow on the toric graph of $X$. This gives an orientation to each edge of the toric diagram, to be along the flow. Moreover, this divides the trivalent vertices into one of two types, depending on whether one or two edges are incoming at the vertex. The two vertices both follow from the refined Chern-Simons theory (these are also the same as the pair of vertices needed in \cite{CIV} and \cite{AK}). The gluing of the vertices is dictated by their origin as refined topological string amplitudes with branes. We will show in detail how our ${\mathbb P}^2$ results fit in this framework. It is easy to show that all of the results obtained previously in \cite{CIV, AK} do so as well.  We will leave developing this into a fully general formalism for future work \cite{ASO}.

The paper is organized as follows. In section 2 we review aspects of refined Chern-Simons theory we will need. We also discuss  D-branes in the refined topological string more generally. In particular, we derive the refined Ooguri-Vafa operators, which are needed to relate the knot invariants in Chern-Simons theory to refined topological string amplitudes with branes. In section 3 we review the large $N$ duality of the refined Chern-Simons theory on $S^3$. As an application of our results from section 2, we show that large $N$ duality of the refined topological string explains a "mirror symmetry" that relates invariants of a knot colored by a Young diagram $R$ and its transpose $R^T$, which was recently discussed in \cite{GukovStosic}. In section 4, we
present the derivation of the refined topological vertex from refined Chern-Simons theory. In section 5, we explain how the work of \cite{AMV} generalizes to the refined topological string. In section 6 we discuss in detail the example related to local ${\mathbb P}^2$. In section 7 we discuss the refined topological vertex formalism. We end with appendices containing necessary definitions and derivations.

\section{Refined Topological String and Chern-Simons Theory}

Refined A-model topological string partition function on a Calabi-Yau $X$is defined as the index of $M$ theory \cite{Vafa-Iqbal, QF, Dijkgraaf2006} on
$$(X\times TN \times S^1)_{q,t}
$$
where $TN$ is the Taub-Nut space, with complex coordinates $z_1$ and $z_2$.
The subscript denotes the $\Omega$-background \cite{N,NO}. Namely, as we go around the $S^1$, the complex coordinates rotate by
\beq\label{Omega}
(z_1, z_2) \rightarrow (q z_1, t^{-1} z_2).
\eeq
Moreover, to preserve supersymmetry, this has to be accompanied by an $R$-symmetry twist.
The M-theory partition function is the index \cite{NO, NO2},

\beq\label{indexa}
Z(M)=   {\rm Tr}\, (-1)^{F}\, q^{S_1-S_R}\, t^{S_R-{S}_2}.
\eeq
of the resulting theory on $TN\times S^1$. Above $S_1$ and $S_2$ generate rotations around the two complex planes, and $S_R$ is the R-symmetry generator.
For $q=t$, the M-theory partition function is the same as the topological A-model string partition function on $X$, where $q$ is related to the topological string coupling by $q=e^{g_s}$. For $q\neq t$ the M-theory partition function defines the partition function of the refined topological string.

In the ordinary topological A-model string, in addition to closed strings, the theory has D-branes wrapping Lagrangian submanifolds $L$ of $X$. In the M-theory formulation, these correspond to M5 branes that wrap
$$
L \times {\mathbb C} \times S^1
$$
where ${\mathbb C}$ is either the $z_1$ or the $z_2$ plane in \eqref{Omega}. We will denote the two complex planes by
${\mathbb C}_q$ and ${\mathbb C}_{\bar t}$, respectively, to help us recall how they are rotated differently by the $\Omega$-background.
The M-theory partition function, in the presence of these branes defines the partition function of the refined open plus closed topological string.

The M-theory partition function will depend on the plane the M5 branes wrap. The M5 branes wrapping the $z_1$ plane, or the $z_2$ plane, and the same lagrangian $L$,  gives rise to two {\it distinct} branes on $L$ in the refined topological string on the Calabi-Yau, \cite{DGH, ACDKV}. To distinguish them, we will call them the refined {\it ${q}$-branes} and the  {\it ${\bar t}$-branes}.

\begin{itemize}
\item[]{ $\qquad\qquad$ M5 brane on  $L \times {\mathbb C}_q \times S^1$  $\qquad \rightarrow\qquad $  {\it ${q}$-brane} on $L$}

\item[]{${\qquad}\qquad$ M5 brane  on $L \times {\mathbb C}_{{\bar t}} \times S^1$ $\qquad \rightarrow \qquad$  {\it $\bar{t}$-brane} on $L$}
\end{itemize}

The bar over $t$ is there to remind us that in the ordinary topological string, the $q$ and ${\bar t}$ branes become the topological branes and anti-branes \cite{ACDKV}.  Note however that in M-theory, they preserve the same supersymmetries. To understand the theory fully, one has to study both of these branes.\footnote{ In the B-model context,
 the $q$- and ${\bar t}$-branes become the degenerate operators of Liouville \cite{Liouville, DGH, ACDKV}. From the four dimensional perspective, they are the surface operators in the four dimensional gauge theory. For other work on branes in the refined topological string see \cite{refined}.}

\subsection{Refined Chern-Simons Theory}

The string field theory on $N$ topological D-branes wrapping a three manifold $M$ inside $T^*M$, is the $SU(N)$ Chern-Simons theory on $M$, where the level $k$ of Chern-Simons is related to $q$, by $q=e^{2\pi i \over k+N}$.  Refined topological string and M-theory allow one to define a refinement of Chern-Simons theory.

In \cite{AS}, we formulated the refined Chern-Simons theory on a three-manifold $M$ as partition function of $N$ M5 branes on
wrapping
$$
M \times {\mathbb C} \times S^1
$$
in M-theory on
$$(T^*M\times TN \times S^1)_{q,t}.
$$
The refined Chern-Simons theory partition function is, per definition, the index of the theory on M5 branes
\beq\label{indexb}
Z(M)=   {\rm Tr}\, (-1)^{F}\, q^{S_1-S_R}\, t^{S_R-{S}_2}.
\eeq
Here $S_1$ and $S_2$ are generators of rotations around $z_1$ and $z_2$, and $S_R$ is the R-symmetry, as before.
This R-symmetry exists when $M$ is a Seifert manifold. This means that $M$ is an $S^1$ fibration over a Riemann surface. The $U(1)_R$ symmetry corresponds to a rotation in the two-plane sub bundle of $T^*M$ consisting of those cotangent fibers $T^*M$ that are co-normal to the generator of the rotation along the Seifert fiber.

The partition function of the M5 brane theory depends on which two-plane in the TN space the M5 branes wrap, i.e. whether we have $q$-branes of the ${\bar t}$ branes wrapping $M$ in M-theory.  So we get two distinct refined Chern-Simons theories, we could call, loosely, $SU(N)_q$ or $SU(N)_{{\bar t}}$. Thus, we get two distinct refinements of the ordinary Chern-Simons theory.
Of course, the partition functions -- in the absence of knots or links, are simply exchanged by a symmetry that takes $(q,t)$ to $(t^{-1}, q^{-1})$.

$$
SU(N)_q \qquad \stackrel{(q,t) \rightarrow ({t}^{-1}, {q}^{-1})}{\xrightarrow{\hspace*{1.7 cm}} }\qquad SU(N)_{\bar t}
$$
Note the subscript {\it does not} directly relate to the level $k$ of the theory.

In the ordinary Chern-Simons case, the $SU(N)$ Chern-Simons theory partition function on a Seifert manifold is computable, by cutting and gluing,  in terms of $S$ and the $T$ matrices, acting on the Hilbert space of the theory on ${T^2}$. In the refined case,  $S$ and $T$ matrices depend on both $q$ and $t$.
But, the Hilbert space of the theory remains the same.
In particular, in $SU(N)_q$ refined Chern-Simons theory, at $q = e^{2 \pi i\over k+\beta N}$, $t=e^{2 \pi i \beta \over k+\beta N}$, for any $\beta$, the Hilbert space is finite dimensional and labeled by representations of $SU(N)_k$ (see \cite{AS, AS2} for more details). In particular, this is independent of $\beta$.
In \cite{AS} the $S$ and the $T$ matrices of the refined Chern-Simons theory were explicitly computed. The partition functions of the $SU(N)_q$ and $SU(N)_{\bar t}$ theory on Seifert manifolds $M$ give rise to new invariants of these manifolds.

\subsection{Knot invariants and branes}

Consider the index of M-theory on $(T^*M \times TN \times S^1)_{q,t}$ with M5 branes on $M$ as before, leading to refined Chern-Simons on $M$. Now we introduce additional M5 branes which we choose to wrap
$$
L_K \times {\mathbb C} \times S^1.
$$
Here, $L_K$ is a
a Lagrangian in $T^*M$,
with the property that it intersects $M$ on a knot $K$
$$
L_K \cap M = K.
$$
$L_K$ is obtained from the knot $K$ in $M$ by a co-normal bundle construction, as explained in \cite{OV}: we consider a knot together with a 2-plane bundle in the fibers of $T^*M$ over it. The fibers of this bundle consist of cotangent vectors orthogonal to the knot. It is clear from the construction that the R-symmetry we need preserves the Lagrangian $L_K$.
In this background, we will end up studying invariants of knot $K$ in the refined Chern-Simons theory. We would like to understand precisely what combination of knot invariants the M-theory index computes.

In the presence of additional branes on $L_K$ the theory gets a new sector, corresponding to M2 branes with ends on both $L_K$ and $M$.  The M2 branes are charged under the fields on both stacks of M5 branes. Consider the contribution of these M2 branes to the index \eqref{indexb} first, where we view the theory on $L_K$ and $M$ as just providing a background. In computing the index, it is natural to turn on fugacities $U$, and $V$, to keep track of these. $U$ and $V$ are the holonomies of the gauge fields on $M$ and $L_K$ around the knot. The gauge field on $M$ arizes by taking the period
of the M5 brane world-volume two-form $B$ along the thermal $S^1$. Their contributions will have an effect of inducing knot observables to the refined Chern-Simons theory. The natural question is what is the corresponding observable.
This observable provides a translation between knot invariants of refined Chern-Simons theory and indices in M-theory.

To understand which observable one gets, one may zoom in on the intersection of $M$ and $L_K$, since the BPS states of M2 branes
 will be localized there.  $M$ and $L_K$ intersect along an $S^1$, which is the copy of the knot.
 Note that, by construction, $L_K$ has one real dimensional moduli space which actually allows us to lift it off $M$. This is because $L_K$ is topologically ${\mathbb R}^2\times S^1$ and by a theorem of MacLean (see \cite{AV12} for a recent discussion) that says that the moduli space of a Lagrangian has dimension $b_1$.  The local geometry of the Calabi-Yau near the intersection is that of ${\mathbb C}^*\times {\mathbb C}^2$, where ${\mathbb C}^*$ is a cylinder that contains the $S^1$. The $M$ and $L_K$ in this neighborhood simply look like two Lagrangians of topology ${\mathbb R}^2 \times S^1$, each wrapping an $S^1$ in ${\mathbb C}^*$. The one real dimensional moduli space of geometric deformations of $L_K$ is parameterized by sliding the corresponding $S^1$ along the cylinder.  The branes intersect only if their positions on the cylinder coincide. Let $\Lambda$ be the Kahler parameter of the annulus, the section of ${\mathbb C}^*$ between the branes. Then, $\log(\Lambda)$ is the mass of the M2 branes whose contributions we want to evaluate. We will fix the branes on $M$ to be $q$-branes, and then it remains to consider either the $q$-branes or the ${\bar t}$-branes along $L_K$. The other two cases are related to this by the symmetry of the theory that takes $(q,t)$ to $(t^{-1}, q^{-1})$.

Consider first the case branes on $M$ are $q$ branes and branes on $L_K$ are ${\bar t}$-branes.
Let us denote by
$${\cal O}_{q{\bar t}}(\Lambda; U, V)
$$
the contribution of M2 branes to the partition function. In the {\it unrefined case}, this operator was computed in \cite{AMV}, following  \cite{OV}, where one found that the partition function is simply
\begin{equation}\label{qtu}
{\cal O}_{q\bar{t}}(U, V)_{q=t} = {{\rm det} (1- U\otimes V^{-1})}.
\end{equation}
This is computed as an annulus diagram in the open topological string, corresponding to integrating out a single bifundamental particle of mass $\log(\Lambda)$, and charged as a bifundamental under the gauge fields on $M$ and $L_K$. Moreover, the particle turns out to be fermionic. We will now argue that in the refined case the answer does not in fact change at all!

From the M-theory perspective, the branes on $M$ and $L_K$ intersect over points on the Taub-Nut space. Since the branes are just supported on points in $TN$, there should be no effect at all: we still get a {\it single}, fermionic BPS particle with no spin or angular momentum on $z_1$ or $z_2$ planes, since it is localized to live at the origin of both of these two planes , and cannot spin. We simply now need to know its charge $s_R$ under the $U(1)_R$ R-symmetry. If we take $s_R$ be $1/2$, in the refined case we get simply
\begin{equation}\label{qta}
{\cal O}_{q\bar{ t}}(U, V) = {{\rm det} (1- v^{-1} U\otimes V^{-1})}.
\end{equation}
where $v=(q/t)^{1/2}$. Before we go on, note that  \eqref{qtu} had a natural expansion in terms Wilson-loop operators. Namely

$$
{\cal O}_{q{\bar t}}(\Lambda; U, V)=\sum_R  (- v^{-1} \Lambda)^{|R|}  \; {\rm Tr}_R U \;{\rm Tr}_{R^T}V^{-1}
$$
here
$$
{\rm Tr}_R\ U
$$
is the holonomy $U$ of the gauge field on $M$ along the knot $K$ in representation $R$, the sum runs over all Young diagrams, and $|R|$ is the number of boxes in the Young diagram corresponding to $R$. This is the Wilson-loop operator of ordinary Chern-Simons theory. Thus, the knowledge of ${\cal O}_{q{\bar t}}(\Lambda; U, V)$
allows us to translate between M-theory, or topological string observables, and Chern-Simons theory.

In the refined $SU(N)_q$ Chern-Simons theory, the operator inserting the Wilson loop in representation ${\mathbb R}$ is \cite{AS, AS2} on $q$-branes is no longer ${\rm Tr}_R U$ but

$$
{\rm Tr}_R U \qquad  {\stackrel{q\neq t }{\longrightarrow}} \qquad M_R(U; q,t)
$$
where
 $M_R(U; q,t)$ is the Macdonald function in representation $R$ \cite{Macdonald}.
 If we consider ${\bar t}$-branes instead, the ordinary traces get traded for $M_R(U;{t}^{-1}, { q}^{-1})$,

 $$
{\rm Tr}_RU \qquad  {\stackrel{q\neq t }{\longrightarrow}}\qquad   M_R(U;  {t}^{-1}, {q}^{-1})
$$
 The fact that there are two different kinds of branes naturally mirrors the fact that there are two different ways to deform Wilson loop operators of the unrefined theory:
 either to $M_R(U;q,t)$ for $q$-branes, or to $M_R(U; t^{-1},q^{-1})$, in the case of the $t$ branes.
At $q=t$, Macdonald polynomials become independent of $q$ and reduce back to ordinary traces.
Macdonald polynomials $M_R(U;q,t)$ can be expanded in terms of finite sums of $ {\rm Tr}_QU $ with coefficients that depend on $q$ and $t$. Thus, in a sense, this is just a convenient basis, in terms of which the $S$ and the $T$ matrices have a particularly simple form \cite{AS, AS2}.

To translate ${\cal O}_{q{\bar t}}$ into knot observables on $M$, we need to be able to expand this in terms of Wilson loop operators natural for ${\it both}$ the $q$-branes on $M$ and the ${\bar t}$-branes on $L_K$. These, as we just explained, are no longer the ordinary traces, but the appropriate Macdonald polynomials. Magically, string theory seems to know about these, and we get a very natural expansion:
In fact, all that happens is that the unrefined Wilson-loop operators get replaced by their refined counterparts:

\begin{equation}\label{qt}
{\cal O}_{q{\bar t}}(\Lambda; U, V) = \sum_R  (-v^{-1} \Lambda)^{|R|} \; M_R(U;q,t) \;M_{R^T}(V^{-1}; {t}^{-1}, {q}^{-1}).
\end{equation}

Now consider the case when both the branes on $M$ and $L_K$ are  $q$-branes, and we have $N$ of branes on $M$ and some number of branes on $L_K$.  We will denote by
 $$
 {\cal O}_{qq}(\Lambda; U, V)
 $$
 the effect on the refined Chern-Simons theory on $M$ of the M2 branes stretching between $M$ and $L_K$. This case will turn out to be somewhat subtle, as we now explain. Consider first the case when the gauge fields on $M$ and $L_K$ are non-dynamical, and we can treat them as providing the background.  Zooming in at the ${\mathbb C}^*\times {\mathbb C}^2$ neighborhood of the intersection, the theory on M5 branes has ${\cal N}=4$ supersymmetry in three dimensions, since the branes coincide on ${\mathbb C}_1\times S^1$, and the Calabi-Yau is flat space. Then, the M2 branes give a hypermultiplet living on ${\mathbb C}_1\times S^1$. The hypermultiplet, from the perspective of ${\cal N}=2$ supersymmetry, consists of two chiral multiplets $Q$ and ${\tilde Q}$ transforming in $(N, {\bar M})$ and $(M,{\bar N})$ representations, respectively. To determine their contributions to the index, we need to know their charges under $S_1$, $S_2$ and $S_R$.  It is easily seen that we can assign $(s_1,s_2,s_R) = (0,0,-1)$ for $Q$, and take ${\tilde Q}$ to be neutral under all.\footnote{
  $S_1$ generates a Lorentz symmetry, and $Q$ and ${\tilde Q}$ are neutral under it. To find the charges under $S_2$ and $S_R$, we can proceed as follows. The underlying ${\cal N}=4$ supersymmetry implies that the hypermultiplets couple to the background adjoint chiral fields $\phi, \phi'$ on the two branes via a superpotential interaction, $\int d^2\theta {\rm Tr} (Q\phi {\tilde Q})-( {\tilde Q}{\tilde \phi} Q)$. The superpotential has to be neutral under the $S_2,S_R$, for these to be symmetries of the theory. The $S_2$ and $S_R$ both generate $R-$symmetries (with their difference being a global symmetry), so $d^2\theta$ term has charge $+1$ under both. The adjoint chiral fields corresponds to the positions of the M5 branes in ${\mathbb C}_{t^{-1}}$ direction, and thus has charge $-1$ under $S_2$, and is naturally neutral under $S_R$ that corresponds to rotations in directions that this field is not sensitive to. We can take $Q$ to have charges $0$ and $-1$ and ${\tilde Q}$ charges $0$ under $S_2$ and $S_R$ to satisfy the requirement of neutrality.}  From this, we can read off their contributions to the index of the theory on ${\mathbb C}\times S^1$:\footnote{By computing the partition function of the two 3d chiral fields on ${{\mathbb C}_1\times S^1}$, or by considering a gas of spinning M2 brane particles on ${\mathbb C}_1$, $Q$ and $\tilde Q$ give %
$
\prod_{n=0}^{\infty}\det(1-q^{n+1} t^{-1}\;\Lambda^{-1} \; U^{-1} \otimes  V)^{-1} \det(1- q^n  \;\Lambda\; U\otimes V^{-1})^{-1}.
$
This can be rewritten, using the property of the quantum dilogarithm function $\Phi(x,q)=\prod_{n=0}(1-q^{n+1/2} x)$ that $\Phi(x,q) = 1/\Phi(x^{-1}, q)$.} as
$$
\prod_{n=0}^{\infty}\frac{\det(1-q^n  t \;\Lambda\;  U\otimes V^{-1})}{\det(1- q^n \;\Lambda\; U\otimes  V^{-1})},
$$
Note that, for $q=t$, this agrees with the unrefined answer,
\begin{equation}
{\cal O}_{qq}(U, V; \Lambda)_{q=t} = {\rm det} (1- \;\Lambda\;  U\otimes V^{-1})^{-1}
\end{equation}
computed in \cite{OV} by different methods. This suggests that we should identify

\begin{equation}
{\cal O}_{qq}(U, V; \Lambda)=\prod_{n=0}^{\infty}\frac{\det(1-q^n  t \;\Lambda\;  U\otimes V^{-1})}{\det(1- q^n \;\Lambda\; U\otimes  V^{-1})},
\end{equation}
This has a simple expansion in terms of holonomy operators of $q$-branes on $M$ and $L_K$, as

\begin{equation}\label{qq}
{\cal O}_{q{ q}}(\Lambda; U, V)=\sum_R  {\Lambda^{|R|}}  \; M_R(U; q,t) \;M_R(V^{-1}; q,t)/g_R
\end{equation}
where $g_R$ depends on $q$ and $t$ and is defined in the appendix A.

The subtlety we alluded to is the following. Per their definition, the operators  ${\cal O}_{q{\bar t}}$, ${\cal O}_{qq}$ describe the {\it effect} of the M2 branes stretched between $M$ and $L_K$ on Chern-Simons theory on $M$. Namely, computing their expectation values one obtains the M-theory index of $N$ $q$-branes on the $M$ and some number of $q$-branes on  $L_K$. While the index is counting BPS states, this does not mean that we can evaluate ${\cal O}_{q{\bar t}}$ or ${\cal O}_{qq}$ by simply counting BPS states of M2 branes between $M$ and $L_K$. These are two a priori unrelated statements. They will be related if we decouple the modes on $M$ and $L_K$, or if the coupling between the M2 branes and M5 brane degrees of freedom is minimal. The latter was always the case in the unrefined theory, as can be seen from the arguments in \cite{OV}. Fortunately, this is often the case in the refined theory as well. The only case that appears subtle is the ${q}$-${q}$ system. Moreover, if either the branes on $L_K$ are non-compact, or if there is an infinite number of them, then the naive computation of ${\cal O}_{qq}$ in terms of counting BPS states applies. When the naive computation fails, the effect is quite simple and this also makes it fairly transparent  what makes the $q$-$q$ (or ${\bar t}$-$\bar t$) system special. When $L_K$ is compact, and the branes on it are dynamical, the operator that replaces \eqref{qq}  turns out to be

\begin{equation}\label{qqs}
{\cal O}^*_{q{ q}}(\Lambda; U, V)=\sum_R  {\Lambda^{|R|}}  \; M_R(U; q,t) \;M_R(V^{-1}; q,t)/G_R
\end{equation}
 Here $G_R$ is the Macdonald metric for the branes on $L_K$. This depends $explicitly$ on the number $N_L$ of branes on $L_K$. This dependence on the rank means that the gauge fields on $L_K$ can not be ignored in answering our question --  this contradicts the decoupling assumption that went into \eqref{qq}. Moreover, $G_R$ reduces to $g_R$ only when number of branes on $L_K$ becomes infinite,
 $$
 \lim_{N_L \rightarrow \infty} G_R = g_R.
 $$
 We have denoted the $qq$-operator in this more subtle case by ${\cal O}^*_{qq}$, to indicate this fine-print. One way to see that ${\cal O}^*_{qq}$ is correct is to specialize to the case we take the number of branes on $M$ and $L_K$ to be equal, i.e. $N=N_L$. Then, it is easy to see that, at $\Lambda=1$, so the branes intersect, we can in fact glue the branes together over the $S^1$. The partition function of glued branes and unglued ones has to be exactly the same, as we are computing an index, which is invariant under all deformations (at least as long as no states run off to infinity. The gluing can be made into a smooth, local, operation so the index as to be invariant under it). The gluing corresponds to cutting out a solid torus neighborhood of the $S^1$ from each Lagrangian, and inserting the Chern-Simons propagator $\sum_R |R\rangle \langle R|/G_R$. In the holonomy basis, this is nothing but the  operator ${\cal O}^*_{qq}(U,V; 1)$ in this case. For most of the paper, all we will need will be the simpler operator ${\cal O}_{qq}$. However, in sections 5 and 6, all of our branes will be compact, and we will need to replace ${\cal O}_{qq}$ with ${\cal O}^*_{qq}$.

 The operators ${\cal O}_{qq}$ and ${\cal O}_{q{\bar t}}$, translate the M-theory index to computations in the refined $SU(N)_q$ Chern-Simons theory on $M$. The M-theory index with $N$ $q$-branes on a compact three manifold $M$ is computed by taking the expectation value, in the refined $SU(N)_q$ Chern-Simons theory on $M$, of either ${\cal O}_{qq}$ or ${\cal O}_{q{\bar t}}$, depending on whether we have the $q$-branes or the ${\bar t}$-branes on $L_K$.
Since $M_R(U; q,t)$ is the operator inserting a Wilson loop in the refined $SU(N)_q$ Chern-Simons theory, if we denote by
$$
 Z_{SU(N)_q}(M, K; R)=\langle M_R(U; q,t) \rangle_{SU(N)_q}
$$
 the refined Chern-Simons partition function with on Wilson loop in representation $R$ along the knot $K$, the M-theory index becomes
either
\begin{equation}\label{Zqq}
Z(M, K; V)_{qq} =  \sum_R    Z_{SU(N)_q}(M, K;R)/G_R \; M_R(V^{-1}; q,t)
\end{equation}
with $q$-branes on $L_K$ or
\begin{equation}\label{Zqt}
Z(M, K; V)_{qt} =  \sum_R  (-1)^{|R|} Z_{SU(N)_q}(M, K;R)\; M_{R^T}(V^{-1}; q,t)
\end{equation}
with branes. For simplicity we have taken $M$ and $L_K$ to intersect, so $\Lambda=1$ here.
Since $L_K$ is a non-compact Lagrangian, the holonomy $V$ at its infinity is a parameter -- it remains as a fugacity, keeping track of M2 brane charges. Note that the partition functions (\ref{Zqq}),(\ref{Zqt}) are different, as $q$- and ${\bar t}$-branes interact differently. The explicit expression for
$
Z_{SU(N)_q}(M, K; R)
$
can always be written in terms of the $S$ and $T$ matrices of the theory in a universal way, as long as $M$ is a Seifert manifold, and $K$ a Seifert knot. The details of the theory only enter is the particular representation of $SL(2,Z).$
Turning this around, while the partition functions with different kinds of branes on $L_K$ are not the same, they contain identical information -- knowing either of \eqref{Zqq}\eqref{Zqt}, we can reconstruct the other.

\section{Branes and Large N transitions}

\begin{figure}[ht]
  \begin{center}
    \includegraphics[width=4in]{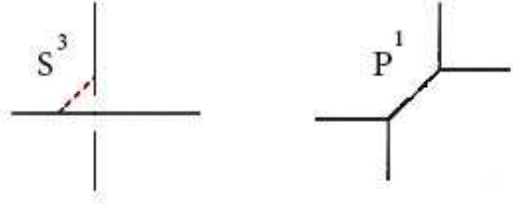}
  \end{center}
    \caption{The geometric transition relating $T^*S^3$ and $Y={\cal O}(-1)\oplus{\cal O}(-1)\rightarrow {\mathbb P^1}$.}
    \label{Transition}
\end{figure}

The ordinary $SU(N)$ Chern-Simons theory on $S^3$ has a large $N$ dual which is the topological string on ${\cal O}(-1)\oplus {\cal O}(-1) \rightarrow {\mathbb P}^1$. Recall, from section 2, that $SU(N)$ Chern-Simons theory on $S^3$ is the same as the open topological string on
$$X=T^*S^3,$$
with $N$ D-branes on the $S^3$.
Gopakumar and Vafa showed this has a large $N$ dual, the ordinary topological string theory on
$$
{Y} = {\cal O}(-1) \oplus {\cal O}(-1) \rightarrow {\bf P}^1.
$$
The duality is a large $N$ duality in the sense of 't Hooft \cite{thooft}. The duality in this case also has a beautiful geometric interpretation: it is a geometric transition that shrinks the $S^3$ and  grows the ${\bf P^1}$ at the apex of the conifold, thereby taking $X$ to $Y$, see figure (\ref{Transition}).

The rank $N$ of the gauge theory is related to the area of the ${\bf P}^1$ in $Y$ by
$Q= e^{-Area({{\bf P}^1})}$ by
\begin{equation}
\label{eq-area}
Q = q^N
\end{equation}
where $q = e^{g_s}.$  The topological string coupling $g_s$ is the same on both sides -- it is related to the level $k$ in Chern-Simons theory by
$g_s = {2\pi i \over k+N}$.  The duality has been checked, at the level of partition functions,  to all orders in the $1/N$ expansion \cite{GV}. An extension of this, where one replaces the $S^3$ by lens spaces, was studied in \cite{AKMV}.
Translated to the M-theory language, large $N$ duality states that the index of the theory before and after the transition are the same. This does not imply that the full theories are the same, but only the index.

It is natural to ask whether this duality extends to the refined case, when we consider the general $q\neq t$ index on both sides? It was shown in \cite{AS} that the partition function of $SU(N)_q$ refined Chern-Simons theory on $S^3$  indeed equals the partition function of the refined topological string on $Y$,

$$
Z_{SU(N)_q}(S^3; q,t) =  Z_Y(Q; q,t)
$$
where
\begin{equation}
\label{eq-area-ref}
Q = t^N (t/q)^{1/2}
\end{equation}
The partition function of the  theory on $S^3$ is computed by the vacuum matrix element of the refined $SU(N)_q$ Chern-Simons theory\footnote{The overall normalizations are slightly ambiguous. The $M$ theory partition function with $N$ $q$-branes on $S^3$ can has a large $N$ limit that gives the conifold with either $Q=t^{N} \sqrt{q/t}$ or $Q=t^{N} \sqrt{t/q}$. The first arises from ${S^{0}}_0$, the second from $S_{00}$. They can both be physical, depending on slightly different details of the setup. Distinguishing these very precisely for the most part will be beyond the scope of this paper.}

$$
Z_{SU(N)_q}(S^3; q,t)= \langle 1\rangle_{SU(N)_q}= {S}_{00}(N; q,t)
$$
where
\begin{align}
\label{nom}
S_{00}(N: q,t)= {{i}^{N(N-1)/2}\over N^{1\over 2}(k+\beta N)^{N-1\over 2}}
\; \prod_{m=0}^{\infty}\prod_{\alpha>0}\frac{(q^{-m/2} t^{-(\alpha,  \rho)/2} - q^{m/2} t^{(\alpha,  \rho)/2}) }{
(q^{-m/2} t^{-(\alpha,  \rho)/2-1/2} - q^{m/2} t^{(\alpha,  \rho)/2+1/2}) \;}\;
\end{align}
The large $N$ duality implies that this is equal to the partition function after the transition

\begin{align}
\label{refc}
Z_Y(Q; q,t) = \exp\Bigl(-\sum_{n=0}^{\infty} {Q^n \over n (q^{n/2} - q^{-n/2})(t^{n/2} - t^{-n/2})}\Bigr).
\end{align}
It is easy to check, \cite{AS}, that this is indeed the case, up to non-perturbative terms, of order $e^{-1/N}$, provided $Q$ is as in equation \ref{eq-area-ref}.

When we include knots in the theory, we consider the additional branes on $L_K$. Large $N$ duality is a geometric transition, and the Lagrangian $L_K$, being non-compact, simply gets pushed through the transition, to $L_K$, together with branes on it \cite{OV}.
In particular, the branes on $L_K$ do not change.

The large $N$ duality implies that the Chern-Simons knot invariants -- corresponding to say $q$-branes on $L_K$:
\begin{align}
Z_{SU(N)_q}(S^3, K; V) &=  \langle {\cal O}_{qq} \rangle_{SU(N)_q} \cr& =\sum_R    Z_{SU(N)_q}(S^3, K, N) \,\, M_R(V^{-1}; q,t)/g_R
\end{align}
compute the partition function of $q$-branes wrapping the lagrangian $L_K$ after the transition on $Y$. In particular, the partition function of the branes on $Y$ should simply be given by rewriting $Z_{SU(N)_q}(S^3, K; V)$ to absorb the $N$-dependence in $Q= t^N (t/q)^{1/2}$.
\subsection{A "Mirror Symmetry" of Knot Invariants at Large $N$}

Before the transition, the partition function of the theory depends sensitively on whether we have $q$-branes on the $S^3$, or the ${\bar t}$-branes.
After the transition, the branes on the $S^3$ disappear, and are replaced by a ${\mathbb P}^1$ of Kahler modulus $Q$. The {\it only} information about what how many, and what kind of branes there were on $S^3$ is in $Q$.   In particular, neither the type of the brane, nor their number has to be the same -- as long as the resulting $Q$ ends up the same.

This implies that, keeping everything else fixed, the theories with $N$ $q$-branes on $S^3$ and ${ N'}$ ${\bar t}$-branes on $S^3$ are the same at large $N$

$$
SU(N)_q \; \longleftrightarrow \; SU({N}')_{\bar t}
$$
theories, where $N$ and ${N}'$ are related by

\beq\label{mirror}
t^N (t/q)^{1/2} = Q = q^{-{N'}} (t/q)^{1/2}
\eeq
or
$$
t^N q^{N'} =1.
$$

This has implications on knot invariants as well. When we add branes on $L_K$,  the interaction of the branes on $L_K$ with branes on $S^3$ will differ sensitively on whether we have $q$-branes or ${\bar t}$-branes on the $S^3$, as the corresponding Ooguri-Vafa operators change, as we explained earlier. But, after the transition only the type of brane on $L_K$ matters, since this is the only brane visible on $Y$ after the transition. We get the same theory on $Y$ with the branes on $L_K$, as long as $N$ and $N'$ are related as in \eqref{mirror}.

Let's consider this in more detail. We can fix the type of brane on $L_K$, to be the $q$-brane say. With $N$ $q$-branes on the $S^3$ the partition function of the theory before the transition is

\begin{equation}\label{Zqqb}
Z(S^3, K; V)_{qq} =  \langle {\cal O}_{qq} \rangle_{SU(N)_q} =\sum_R Z_{SU(N)_q} (S^3, K;R)\; M_R(V^{-1}; q,t)/g_R
\end{equation}
Taking instead the ${N'}$ ${\bar t}$-branes on $S^3$, we get

\begin{equation}\label{Zqtb}
Z(S^3, K; V)_{tq} =  \langle {\cal O}_{{\bar t}q} \rangle_{SU(N')_{\bar t}} =\sum_R (-v)^{|R|}Z_{SU(N')_{\bar t}} \; (S^3, K;R^T) M_{R}(V^{-1}; q,t)
\end{equation}
where ${\cal O}_{qq}$ and ${\cal O}_{{\bar t}q}$ are defined in \eqref{qq} and \eqref{qt}. Note that ${\cal O}_{{\bar t}q}$ and ${\cal O}_{q{\bar t}}$ are the same up to $v\rightarrow v^{-1},$ by analytic continuation.
The large $N$ duality implies that \eqref{Zqqb} and \eqref{Zqtb} are equal, after we absorb the dependence of the two amplitudes on $N, N'$
on $Q$. In particular, extracting the coefficient of $M_R(V; q,t),$
for a fixed representation $R$, we see that

\beq\label{conj}
Z_{SU(N)_q} (S^3, K;R)/g_R \stackrel{fixed \;\; Q}{=} (-v)^{|R|} Z_{SU(N)_{\bar t}} (S^3, K;R^T)
\eeq

As a check, note that this is satisfied for the unknot, colored by an arbitrary representation.  In this case

$$
Z_{SU(N)_q} (S^3, \bigcirc;R)/Z_{SU(N)_q} (S^3) = {S}_{R0}/{S}_{00}(SU(N)_q)= M_R(t^{\rho_N}, q,t)
$$
and similarly
$$
Z_{SU(N')_{\bar t}} (S^3, \bigcirc;R)/Z_{SU(N')_{\bar t}} (S^3) = S_{R0}(SU(N')_{\bar t})= M_R(q^{-\rho_{N'}}; t^{-1},q^{-1})
$$
where $\rho_N$ denotes the Weyl vector of $SU(N)$.
Using identities of Macdonald functions, it is easy to prove that the conjecture \eqref{conj} indeed holds for the unknot.
For more general knots, then, it suffices to consider the normalized knot invariant, where one divides by expectation value of the unknot in the same representation $R$. For totally symmetric or totally antisymmetric representations, the resulting knot invariant is the (reduced) superpolynomial, studied in \cite{Superpoly1, Superpoly2, Superpoly3, GukovStosic, Moscow1, Moscow2},
In general, by conjectures in \cite{AS, AS2},

$$
P(K)_R(q,t,Q) =Z_{SU(N)_q} (S^3, K;R)/Z_{SU(N)_q} (S^3, \bigcirc;R)
$$
is the index on the reduced knot homology theory categorifying the colored HOMFLY polynomial.
The conjecture \eqref{conj} implies
$$
P(K)_R(q,t,Q) =P(K)_{R^T}(t^{-1},q^{-1},Q)
$$
This property of the colored polynomials was called mirror symmetry in \cite{GukovStosic}, for the way it acts on the dimensions of knot homologies. In \cite{GukovStosic} a different explanation for the duality was proposed. Note that in the unrefined case, when $q=t$, this implies a symmetry of the colored HOMFLY
polynomial, that in the reduced case says $P(K)_R(q,Q) =P(K)_{R^T}(q^{-1},Q).$ While the symmetry is present even in the unrefined theory, its most natural explanation is in the refined case, as only then \eqref{mirror} rigorously makes sense.
\section{Refined topological Vertex from Refined Chern-Simons Theory}

\begin{figure}[ht]
\begin{center}
\includegraphics[width=4in]{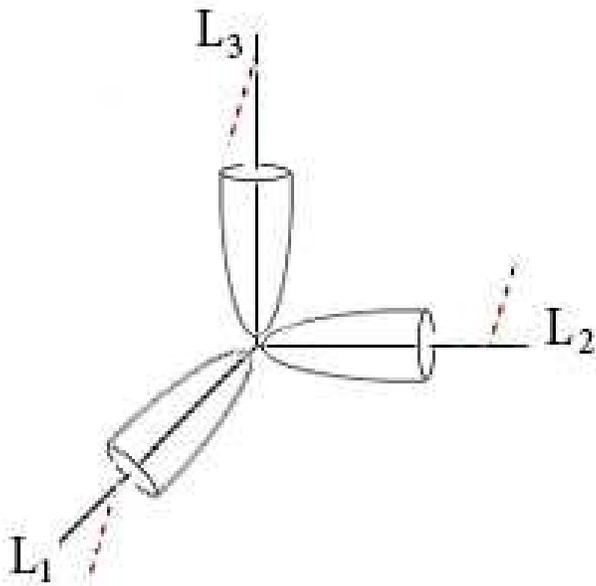}
 \caption{${\mathbb C}^3$ with three stacks of branes on Lagrangians $L_1$, $L_2$, $L_3$.}
 \end{center}
\label{fig-2}
\end{figure}

In \cite{TV}, the ordinary Chern-Simons theory and large $N$ duality was used to derive the topological vertex. In a toric Calabi-Yau, there is a simple class of toric Lagrangian branes $L_i$  \cite{AV01}, which have the topology of ${\mathbb R}^2\times S^1$, and project to lines in the toric base. The topological vertex is the partition function of the topological string on ${\mathbb C}^3$, with formaly infinite number of branes on each of the three toric Lagrangians, as in the figure \ref{fig-2}.

\begin{figure}[ht]
  \begin{center}
    \includegraphics[width=0.5\textwidth]{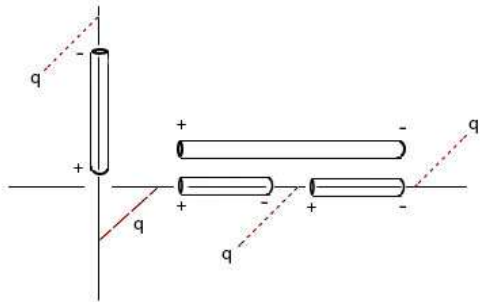}
  \end{center}
    \caption{The $T^*S^3$ with three brane stacks. The M2 branes wrapping with holomorphic annuli are schematically shown.}
        \label{fig-3a}
\end{figure}

\begin{figure}[ht]
  \begin{center}
    \includegraphics[width=0.5\textwidth]{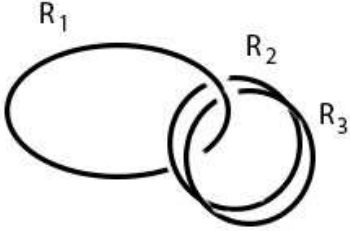}
  \end{center}
  \label{fig-3}
    \caption{A "double" Hopf link, the starting point for derivation of topological vertex. \label{DoubleHopf} }
\end{figure}

Let us briefly sketch the idea of the derivation in \cite{TV}. Consider the large $N$ limit of Chern-Simons theory on $S^3$. This is the conifold  $Y={\cal O}(-1) \oplus {\cal O}(-1) \rightarrow {\bf P}^1$ with the size of the ${\mathbb P}^1$ fixed by the number of branes on $S^3$.  A toric Lagrangian brane on $Y$ corresponds to an unknot in $S^3$ \cite{AV01}, and the three Lagrangian branes on $Y$ correspond to a three-component link in $S^3$, consisting of unknots.   Consider a "double" Hopf-link on the $S^3$, as in the figure \ref{fig-3a}. Namely, we consider a 3-component link consisting of two unknots linking the third, as in the figure (\ref{DoubleHopf}). This corresponds, in the string geometry, to having the $L_1$, $L_2$ Lagrangians along one of the toric legs, and $L_3$ along another.

In the strict $N$ to infinity limit, the size of the ${\mathbb P}^1$ goes to infinity, and $Y$ degenerates into a ${\mathbb C}^3$ -- if, as we take the limit, we zoom in to the neighborhood of one of the vertices. We will use this fact to derive the amplitudes of branes on ${\mathbb C}^3$ from those on $Y$.

\begin{figure}[ht]
  \begin{center}
    \includegraphics[width=0.5\textwidth]{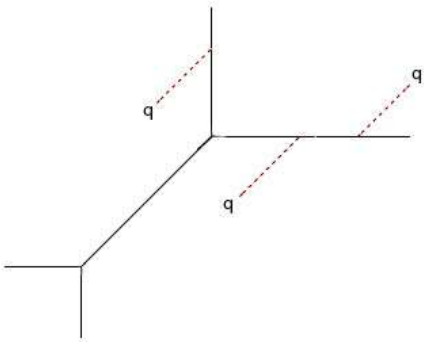}
  \end{center}
    \caption{Branes on the conifold in the large $N$ limit.}
        \label{fig-3f}
\end{figure}

Clearly,  we need to move one of the two stacks of branes to the empty leg. To do this, we have to pass the brane through the vertex. This is a classically singular configuration, where the area of holomorphic disks ending on the Lagrangian vanishes; this area is, classically, proportional to the distance of the brane from the vertex, \cite{AV01}. However the singularity in the classical geometry of the brane moduli space is removed by disk instanton corrections. Summing them up, we obtain the geometry of the Riemann surface of the mirror Calabi-Yau -- in this case, the mirror to ${\mathbb C}^3$. This was recently reviewed in \cite{AV12}. The disk instantons smooth out the geometry of the moduli space: any singularities are in complex codimension one, so they can be avoided. This means that, having obtained the amplitude with two stacks of branes on a single leg, we can simply analytically continue around the singularity, to the configuration we were after.

In this section, we will explain that in the refined context, following analogous steps, we can derive, from the refined Chern-Simons theory and M-theory, the refined topological vertex of \cite{CIV} and \cite{AK}. To begin with, consider $X=T^*S^3$ with $N$ $q$-branes on the $S^3$. Moreover, we consider three lagrangians in this geometry, $L_1$, $L_2$, $L_3$, as in the figure \ref{fig-3a}, with $q$-branes wrapping them. In writing down this amplitude, we made a number of choices, of say $q$-branes versus ${\bar t}$-branes. None of them are essential: the distinction of $q$ branes versus ${\bar t}$-branes on $S^3$ vanishes at large $N$, as we explained in section 3. Changing the type of the branes on $L_i$ also contains no new information: as we will show, having written down any one of the amplitudes, we can obtain from it the others -- the only thing that changes are the wilson loop observables. Moreover, as we will explain later, in defining the amplitude one has to break the symmetries of ${\mathbb C}^3$, so a completely cyclically symmetric vertex does not exist in the refined theory, unlike in the unrefined case in \cite{TV}. This being the case, we will break the symmetries from the outset, and simply pick a convenient choice.

We have to write down the effective contributions of M2 branes ending on the Lagrangians pairwise. These will act as linear combinations of observables in the refined Chern-Simons theory, related to the doubled Hopf-link, in the figure (\ref{fig-3}).
The M2 branes wrapping holomorphic annuli between the branes on $S^3$ and each of the stack of branes on $L_{2,3}$ have the effect of inserting
${\cal O}^{qq}(U, V_i)$ from \eqref{qq}.

$$
{\cal O}_{q{ q}}(\Lambda,  U, V)=\;\prod_{n=0}^{\infty}\frac{{\rm det} (1-q^{n} t\Lambda \; U\otimes V^{-1})}{{\rm det}(1- q^{n}  \; \Lambda \;U \otimes V^{-1})}.
$$
This is because $L_{2,3}$ and the $S^3$ can be made to intersect on $S^1$ at best, all the branes are $q$-branes, and moreover, $L_i$ are noncompact.
Above $U$ is the holonomy on the $q$-branes on the $S^3$, and $V_i$ the holonomy on the branes on $L_i$.   As explained in \cite{TV}, one also has to include the contribution of holomorphic curves that do not end on the $S^3$ - as long as they end on the non-compact Lagrangians, they will contribute to the net amplitude.
The M2 branes wrapping these curves will also contribute to the M-theory index. The only such curves in this geometry come from holomorphic annuli with boundaries
on $L_1$ and $L_2$ -- all other contributions vanish.\footnote{There are of course also M2 branes wrapping annuli beginning and ending on the same stack of $L_i$ branes. These contribute just overall, representation independent factors, that are typically suppressed. See \cite{IntegrableHierarchies}.}  They contribute a factor of ${\cal O}_{qq}(V_1,V_2)$. The boundaries labeled by $+$ and $-$ correspond to  two different orientations of the $S^1$ boundary of the annulus. Changing the $+$ to a $-$ changes the holonomy around the $S^1$ from $V$ to $V^{-1}.$ The fact that operators like ${\cal O}_{qq}$ are not invariant under permutations of branes that would flip the orientation of annuli, implies that we have to keep track of it.  Locally, the relative orientations are fixed, but there is some arbitrariness in choosing the orientations globally. This should be related to the choice of the $U(1)_R$ symmetry,  as we will discuss in section 7.

Finally, consider the contributions of M2 branes with boundaries on the $S^3$ and $L_1$. The configuration of branes is different than that in section $1$. If we were to make the branes intersect, they would not intersect on an $S^1$, as we assumed there. Instead, the branes would simply coincide. It helps to consider the local geometry near the branes, where the Calabi-Yau looks like ${\mathbb C}^*\times {\mathbb C}^2$, where only half of the $S^3$ is visible. Topologically, the half is ${\mathbb R}^2\times S^1$, and this has the same topology as $L_1$. The compactness of the $S^3$ does not affect the operator, but only its expectation value, which we will address later. The index in such a situation has been computed in \cite{AS}, with the result

\begin{equation}\label{qbarq}
{\cal O}_{q\bar{ q}}(\Lambda,  U, V)=\;\prod_{n=0}^{\infty}\frac{{\rm det} (1-q^{n} \Lambda \; U\otimes V^{-1})}{{\rm det}(1- q^{n}  t\; \Lambda \;U \otimes V^{-1})},
\end{equation}
where $\log \Lambda$ is the mass of the M2 branes.  The reason \eqref{qbarq} is just the inverse of \eqref{qq} is that either system can be viewed as a supergroup version of the other \cite{Dijkgraaf-Vafa-DeconstructioPaper}. We will set $\Lambda$'s to $1$ by absorbing them into $V_i$'s for the rest of this section.
%
%

%

Taking all these factors into account, before the transition, the brane configuration in figure (\ref{fig-2}) corresponds to computing the following correlator in the refined Chern-Simons theory:

\beq\label{Ycor}
Z_{SU(N)_q}(S^3, V_1, V_2, V_3)= \langle  {\cal O}_{q{\bar q}}(U, V_1) \;{\cal O}_{qq}(U, V_2)\;{\cal O}_{qq}(U, V_3)\rangle_{SU(N)_q} \; {\cal O}_{q{q}}(V_1, V_2)
\eeq
To compute the amplitude, we have to expand it in link observables of the refined Chern-Simons theory on $S^3$. Recall, from section 2 and appendix A that:
$$
 {\cal O}_{qq}(U, V) = \sum_{Q} M_{Q}(U)\; M_{Q}(V^{-1})/g_Q
 $$
and
$$
 {\cal O}_{q{\bar q}}(U, V) = \sum_{Q} M_{Q}(U) \; {i M}_{Q}(V^{-1})/g_Q
 $$
 where the $i$ operation is defined in appendix $A$.
Since the un-knots colored by $Q_1$ and $Q_2$ are parallel, and linked with the unknot colored by $Q_3$
we have that
\begin{align}
\Big< M_{Q_1}(U) M_{Q_2}(U)  M_{R_3}(U) \Big>_{ SU(N)_q} &= \sum\limits_{P} {\cal N}_{Q_1,Q_2}^{P} \Big< M_P(U)  M_{R_3}(U) \Big>_{SU(N)_q}
\cr
&= \sum\limits_{P} {\cal N}_{Q_1,Q_2}^{P} S_{PR_3}\cr
&=S_{Q_1R_3} S_{Q_2R_3}/S_{0R_3}
\end{align}
where $S$ is the $S$-matrix of the $SU(N)_q$ Chern-Simons theory. Using this, and taking the large $N$ limit, we would obtain the amplitude corresponding to three stacks of branes in ${\cal O}(-1)\oplus {\cal O}(-1)\rightarrow {\mathbb P}^1$ in the figure \ref{fig-3a}. We are interested in branes on ${\mathbb C}^3$, so we will take the $N$ to infinity limit instead. In this limit,

$$
\lim_{N\rightarrow \infty} t^{-N(|R|+|Q|)/2}S_{RQ} = M_{R}(t^{\rho}) M_{Q}(t^{\rho} q^{Q}).
$$

We will absorb the proportionality factor $t^{N(|R|+|Q|)}$ into the definitions of the holonomies $V_i$.
Putting this all together, the partition function of branes on ${\mathbb C}^3$  is\footnote{It should be clear that in this context the operators ${\cal O}_{qq}$, as those appropriate for the infinite number of branes, i.e with $G_R$ replaced with $g_R$.}

\begin{align}
Z_{{\mathbb C}^3}(V_1, V_2, V_3)=\sum\limits_{R_3}
M_{R_3}(t^{\rho})
&{\cal O}_{q{\bar q}}(t^{\rho} q^{R_3}, V_1)
{\cal O}_{qq}(t^{\rho} q^{R_3}, V_2) \cr
 \times&{\cal O}_{{q} {q}}\big( V_1, V_2 \big)  {M}_{R_3}(V_3^{-1})/g_{R_3}
\end{align}

\begin{figure}[ht]
\begin{center}
  \includegraphics[width=3in]{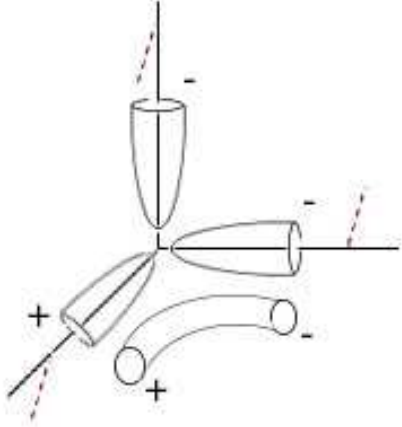}
  \end{center}
    \caption{Analytic continuation in $V_1$ corresponds to moving $L_1$. The orientation of the boundaries is inherited from before the transition.}
    \label{fig-4}
\end{figure}

To get the refined topological vertex, we need to have the three stacks of branes on the different legs. To do this, we will move the branes on the first stack  to the unoccupied leg. This corresponds to analytic continuation in the parameters $V_{1}$ that capture the positions of the branes, from $V_1\gg 1$ to $V_1\ll 1$. Note that in the refined topological string the moduli space of the brane remains exactly the same as in the unrefined case, so we can use the holomorphy of the moduli space, just as in the unrefined case,  to argue there are no phase transitions as we move the branes around. The only factor that needs analytical continuation is ${\cal O}_{q{\bar q}}(t^{\rho} q^{R_3}, V_1)$, as only this factor depends on $V_1^{-1}$; ${\cal O}_{qq}(V_1, V_2)$ depends on $V_1$, and makes sense after we move the brane (see figure \ref{fig-4}). Since ${\cal O}_{q{\bar q}}$ is a product of ratios of quantum dilogarithms,
its analytic continuation corresponds simply to replacing\footnote{The quantum dilogarithm $\Phi(x)= \prod_{n=0}^{\infty}(1-q^{n-1/2}x)$ satisfies  $\Phi(x,q)=1/\Phi(x, q^{-1})$ up to a simple factor that is unimportant for our purposes.  In the present case, the amplitude ${\cal O}_{q\bar q}$ is a product of ratios of quantum dilogarithms. Naively, this is still an infinite product, however, the infinite product can be regularized and rewritten as a finite product of quantum dilogarithms. For details, see appendix B.}

\beq\label{ac}
{\cal O}_{q{\bar q}}(t^{\rho} q^{R_3}, V_1)\qquad \rightarrow \qquad {\cal O}_{q{\bar q}}(t^{-\rho} q^{-R_3}, V^{-1}_1 v^{-2}),
\eeq
as we show in appendix B.
This derivation of the analytic continuation is a slight improvement of the one in \cite{TV}, as it does not rely on the symmetries of the theory.
All in all, the refined topological string amplitude on ${\mathbb C}^3$, with three stacks of branes equals

\begin{align}\label{vert}
{\cal C}(V_1, V_2, V_3)=
\sum\limits_{R_3} M_{R_3}(t^{\rho})
&{\cal O}_{q{\bar q}}(t^{-\rho} q^{-R_3}, V_1^{-1}v^{-2})
{\cal O}_{qq}(t^{\rho} q^{R_3}, V_2)\;
\cr \times &{\cal O}_{{q}, { q}}\big( V_1, V_2 \big)\;{M}_{R_3}(V_3^{-1})/g_{R_3}
\end{align}

Th refined topological vertex amplitudes should correspond to the coefficient of the ${\mathbb C}^3$ partition function, when we expand it in appropriate basis. Since the branes on $L_i$ are $q$-branes, the most natural basis from perspective of refined Chern-Simons theory is the basis of $M_R(U)=M_{R}(U; q,t)$ Macdonald functions, as we explained in section 2. This defines

\beq\label{VAKo}
{\cal C}(V_1, V_2, V_3)=\sum_{R_1,R_2,R_3}\; \;{\cal C}_{R_1 R_2R_3}(q,t)\; M_{R_1}(V_1)/g_{R_1}M_{R_2}(V_2^{-1}) /g_{R_2}M_{R_3}(V_3^{-1})/g_{R_3}.
\eeq
Expanding, we find in the appendix A that

\beq\label{VAK}
{\cal C}_{R_1 R_2 R_3}(q,t) = \sum\limits_{R}
v^{-2 |R|}\ g_R
\ iM_{R_1 / R}(t^{-\rho} q^{-R_3})
\ M_{R_2 / R}(t^{\rho} q^{R_3})
\ M_{R_3}(t^{\rho}).
\eeq
This  is exactly the refined topological vertex amplitude of \cite{AK} (see equation (4.4) of that paper)

$${{\cal C}}_{R_1 R_2 R_3}(q,t) = { C}^{AK}_{R_1 R_2 R_3}(q,t).$$

More precisely, the vertex is the same, up to the change of framing; we have at the outset chosen a different framing for the branes on the second leg. As the vertex has no symmetry, using a cyclically symmetric framing as in \cite{TV} does not buy one anything.\footnote{In \cite{AK}, one had $1$ appearing in the sum, instead of $v^{-2|R|}.$ We believe that is an essentially arbitrary ${\mathbb Z}_2$ choice one gets to make at one place in the theory. See also footnote \ref{subtle}.}

\begin{figure}[ht]
\begin{center}
\includegraphics[width=3.5in]{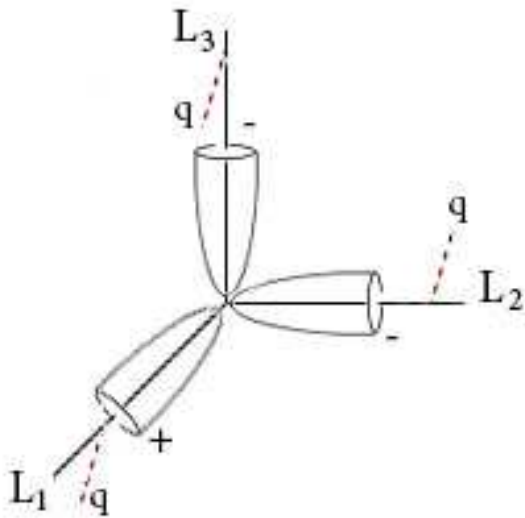}
 \caption{${\mathbb C}^3$ with three stacks of refined $q$-branes on Lagrangians $L_1$, $L_2$, $L_3$.}
 \end{center}
\label{TopVertex}
\end{figure}

In summary have derived the refined vertex of \cite{AK} from the refined $SU(N)_q$ Chern-Simons theory at large $N$. We have shown that the vertex, which was previously used to compute Nekrasov partition functions \cite{N, NO}, has an interpretation as the refined topological string amplitude with branes inserted, analogously to the unrefined case, as anticipated in \cite{CIV}.

\subsection{The Refined Topological Vertex of \cite{CIV}}

There is another, perhaps more famous version of the refined topological vertex, the vertex of \cite{CIV}. The refined vertex of \cite{CIV} has a beautiful combinatorial interpretation in terms of counting boxes. This was recently explained in terms of relating it to the refined Donaldson Thomas theory in 6 dimensions \cite{QF, NO2} (or equivalently, in the IIA formulation of the refined topological string, as counting BPS bound states with a D6 brane \cite{Dijkgraaf2006}).  That vertex {is the same as the
refined topological vertex of} \cite{AK},  up to a change of basis of symmetric functions. Instead of giving an explicit formula for the change of basis, we will give its physical interpretation -- as the change of the refined branes. We will see that the origin of the $q/t$ asymmetry of the box-counting in the vertex is indeed related to the choice of different refined $q,{\bar t}$-branes on two of the legs, as anticipated in \cite{CIV}. But, we will also see that there is nothing exotic about the third leg at all. Let us explain this in some detail.

\begin{figure}
  \begin{center}
    \includegraphics[width=0.5\textwidth]{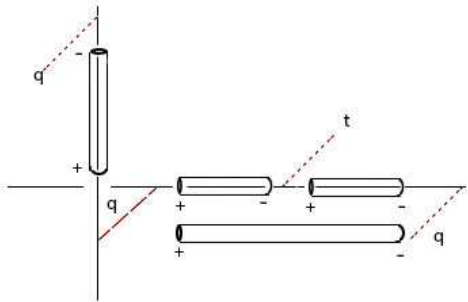}
  \end{center}
    \caption{The topological vertex amplitude with a different choice of brane types. \label{TopVertexDifferent} }
\end{figure}

In choosing which ${\mathbb C}^3$ amplitude to compute, we made some choices.
We could have also changed the configuration of the non-compact branes. Suppose instead we study the configuration of branes on the figure (\ref{TopVertexDifferent}). We changed the branes on $L_1$ from $q$-branes to ${\bar t}$-branes, and moreover we flipped their orientation relative to the plane of the figure. In addition, we flipped the orientation of the branes on $L_2$, keeping them $q$-branes. The operator whose expectation value we compute changes, as ${\bar t}$-branes and $q$-branes interact, and also, changing relative orientation of the $q$ branes also makes them interact differently. Then, the refined topological string partition function is computed by the following correlator:

$$
Z'_{SU(N)_q} (S^3;V_1, V_2, V_3)= \langle  {\cal O}_{q\bar t}(U, V_1) \;{\cal O}_{q{\bar q}}(U, V_2)\;{\cal O}_{qq}(U, V_3)\rangle_{SU(N)_q}\; {\cal O}_{q{\bar t}}(V_1, V_2)
$$

The correlator looks different, but we will now show that the resulting amplitude is closely related to the one we just computed. Repeating the steps of the previous derivation, i.e. taking the $N$ to infinity limit and analytically continuing-- we find the amplitude is equal to

\beq\label{VAKo'}
{\cal C}'(V_1, V_2, V_3)=\sum_{R_1,R_2,R_3} (-1)^{|R_1|}\; {\cal C}_{R_1 R_2R_3}(q,t)\; i{\hat M}_{R_1^T}(V_1) \; iM_{R_2}(V_2^{-1})/g_{R_2} \; M_{R_3}(V_3^{-1})/g_{R_3}.
\eeq

The vertex amplitude that enters, ${\cal C}_{R_1,R_2,R_3}$ is {\it the same as the vertex} amplitude we obtained previously in \eqref{VAK}. The only thing that changes is the basis of symmetric functions containing the holonomies. Let us explain the origin of the change.

Relative to the figure (\ref{fig-3a}), in (\ref{TopVertexDifferent}) the second stack of branes flipped relative to the plane of the paper. In terms of the amplitude, the effect of this is to change

$$
M_{R_2}(V_2^{-1}) \qquad  \rightarrow \qquad iM_{R_2}(V_2^{-1})
$$
The $M_R\rightarrow iM_R$ is the ${\mathbb Z}_2$ involution of symmetric functions defined in the appendix. For the first stack of branes we both flipped the branes and changed the $q$ branes to $t$ branes. Were we to just change the $q$-branes to branes, we would have replaced

$$
 M_{R_1}(V_1)/g_{R_1} \qquad \rightarrow \qquad  (-1)^{|R_1|}\;\hat{{M}}_{R_1^T}(V_1)
$$
where

$$M_{R_1}(V_1)= M_{R_1}(V_1;q,t), \qquad {\hat M}_{R_1}(V_1)= M_{R_1}(V_1;t^{-1},q^{-1})
$$
as the $M_R(V;q,t)$ basis is the natural basis for $q$-branes, and ${M}_R(V;t^{-1}, q^{-1})$ is natural for the ${\bar t}$-branes. Since we flip the branes in addition, corresponds to subsequently applying the involution $i: \, M_R\rightarrow \, iM_R$. It is easy to show that this behavior of the partition function under the two ${\mathbb Z}_2$ actions, one exchanging the $q$ and the $t$ branes, and the other flipping the branes, is a general phenomenon.

Now, let's expand \eqref{VAKo} in the basis of Schur functions as follows

\beq\label{VAKo"}
{\cal C}'(V_1, V_2, V_3)=\sum_{R_1,R_2,R_3} \; {\cal C}^{CIV}_{R_1 R_2R_3}(q,t)\; s_{R_1}(V_1)  \;s_{R_2}(V_2^{-1}) \;M_{R_3}(V_3^{-1})/g_{R_3}.
\eeq

In appendix $D$ we show that the coefficient is the vertex of \cite{CIV}:

\beq\label{VIV}
{{\cal C}^{CIV}}_{R_1 R_2 R_3}(q,t)= \sum\limits_{R} (-v)^{|R|}\;
s_{R_1^T / R}(t^{-\rho} q^{-R_3})\;
s_{R_2 / R}(q^{\rho} t^{R_3^T})\;
 M_{R_3}(t^{\rho})
\eeq
This is exactly the refined topological vertex amplitude of \cite{CIV}, up to change of framing on the second leg.\footnote{\label{subtle}In addition, one needs to replace  the factor $v^{|R|}$ by $v^{-|R|}$. There is a ${\mathbb Z}_2$ ambiguity in defining the theory in both \cite{CIV} and in our approach. In our context, we had to choose the $U(1)_R$ charge in \eqref{qt}; the two choices of $s_R=\pm 1/2$ lead to slightly different amplitudes, but same physics. In \cite{CIV}, one had a choice of how to color the boxes in central slice of the crystal in the box-counting formulation of the vertex, by $q$ or by $t$. With a different choice in \cite{CIV}, our vertices are identical. }
The fact that ${\cal C}^{AK}_{R_1 R_2 R_3}$ and ${\cal C}^{CIV}_{R_1 R_2 R_3}$ are related by a change of basis was shown earlier in \cite{AwataFeigin}, and also discussed in \cite{Iqbal}.

\subsection{The Second Vertex}

There is one choice that we made that may give a new amplitude. This corresponds to a flop of the $S^3$, see figure (\ref{TopVertexConjugate}). The configuration is a-priori different, as, to preserve the same supersymmetry, the branes on the $S^3$ before and after the transition have to be in a different homology class.
To compute the corresponding amplitude, however, we do not need to do a new computation.

\begin{figure}[ht]
  \begin{center}
    \includegraphics[width=0.5\textwidth]{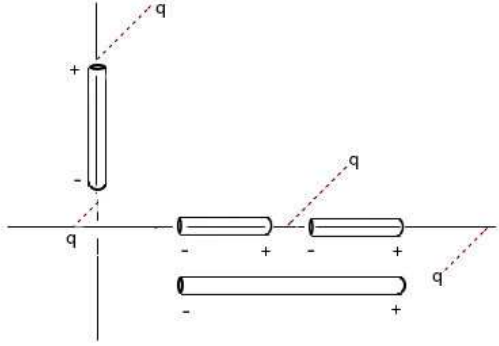}
  \end{center}
    \caption{The brane configuration on $T^*S^3$ leading to the conjugate vertex. \label{fig-3b} }
\end{figure}

Consider the brane configurations in figures (\ref{fig-3a}) and (\ref{fig-3b}). They are related to each other by orientation reversal, where we change the orientation of all the three manifolds at the same time. In Chern-Simons theory, change of the orientation of the three manifold takes $q$ to $q^{-1}$. In the refined Chern-Simons theory we have to reverse both $q$ and $t$, as they are related to the effective couplings of $SU(N)_q$ and $SU(N)_{\bar t}$ Chern-Simons theories, and we could have considered either the $q$ or the $t$ branes.
Thus, to get the amplitude corresponding to figure (\ref{fig-3b}) we need to take

$$
(q,t) \rightarrow (q^{-1}, t^{-1}).
$$

In addition, the fact that the orientation of the non-compact branes changes as well means that we have to send, simultaneously,

$$
\qquad V_i\rightarrow V_i^{-1}.
$$

Applying these operations to (\ref{VAKo}), (\ref{VAK}) implies that the brane configuration in the figure (\ref{fig-3b}) leads to the following vertex:

\begin{figure}[ht]
  \begin{center}
    \includegraphics[width=0.57\textwidth]{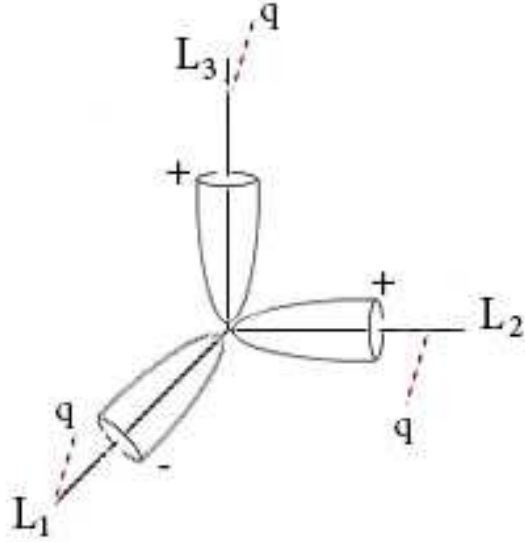}
  \end{center}
    \caption{The second vertex \label{TopVertexConjugate} }
\end{figure}

\beq\label{VAKoc}
{\overline {\cal C}}(V_1, V_2, V_3)=\sum_{R_1,R_2,R_3}\; \;{\overline {\cal C}}_{R_1 R_2R_3}(q,t)\; M_{R_1}(V_1^{-1})/g_{R_1}\;M_{R_2}(V_2)/g_{R_2} \; M_{R_3}(V_3)/g_{R_3}.
\eeq
where

$$
{\overline {\cal C}}_{R_1 R_2 R_3}(q,t)={{\cal C}}_{R_1 R_2 R_3}(q^{-1},t^{-1})
$$
or

\beq\label{VAKc}
{\overline {\cal C}}_{R_1 R_2 R_3}(q,t) = \sum\limits_{R}
 g_R
\ iM_{R_1 / R}(t^{\rho} q^{R_3})
\ M_{R_2 / R}(t^{-\rho} q^{-R_3})
\ M_{R_3}(t^{-\rho}).
\eeq

We used here that ${\overline g_R}= v^{2|R|} g_R(q,t)$, and absorbed the corresponding shift associated with the external representations in the definition of $V_i$'s, thus trading $V_i\rightarrow v^{-2|R|} V_i^{-1}$. This vertex corresponds to the branes in figure (\ref{TopVertexConjugate}).

\section{Large $N$ Duality and More General Geometries}

\begin{figure}
  \begin{center}
    \includegraphics[width=0.5\textwidth]{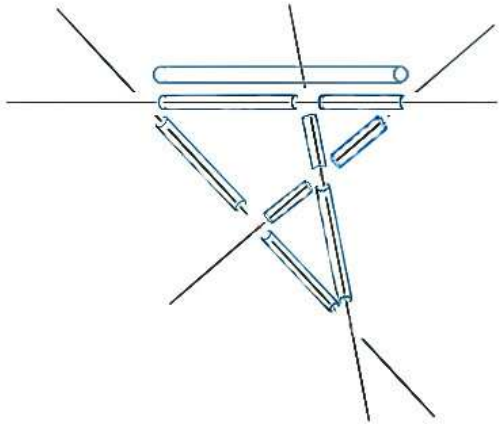}
  \end{center}
\ \ \ \ \ \
  \begin{center}
    \includegraphics[width=0.5\textwidth]{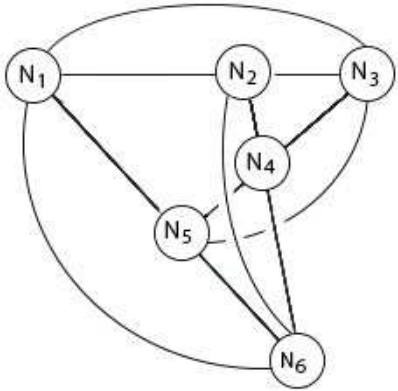}
  \end{center}
    \caption{An example of a geometry for quiver Chern-Simons theory, with holomorphic annuli schematically shown. To the right, the corresponding quiver. \label{GenGeom} }
\end{figure}

As shown in \cite{AMV}, the geometries where several $S^3$'s shrink are related by geometric transitions and large $N$ duality to topological strings on a large class of toric geometries. Namely, when we wrap branes on the $S^3$'s, we get, before the transition,
 quiver Chern-Simons theories, with nodes corresponding to the shrinking $S^3$'s. At large $N$, the theory will have a dual description in terms of a geometry where the $S^3$'s have undergone transitions  and get replaced by ${\mathbb P}^1$'s. Moreover, one can generalize this further to cases where some of the three manifolds are not $S^3$'s but are instead Lens spaces, i.e. orbifolds of $S^3$'s \cite{AKMV}.  This leads to a generalized notion of a quiver were the nodes carry some topological data too, and where different nodes may correspond to different topologies.
In this case, the geometric transitions are slightly more complicated, where instead of ${\mathbb P}^1$'s complex surfaces can open up at large $N$.  It is natural to expect that refined Chern-Simons theory leads similarly, via large $N$ transitions, to refined topological string amplitudes on these geometries. We will show that this is indeed the case, in the examples we have studied.
Quiver Chern-Simons theories arize from local $T^2\times {\mathbb R}$ fibered Calabi-Yau  manifolds with no holomorphic 2-cycles. An example of such a Calabi-Yau is in the figure (\ref{GenGeom}).
The graph $\Gamma$ that captures the geometry of the Calabi-Yau and its singular $T^2$ fibers is a set of straight lines of integer slope
in the ${\mathbb R}^3$ base. There are no closed holomorphic 2-cycles in the geometry, but there are minimal three-cycles. Their geometry is encoded in $\Gamma$ in simple way. In fact, the projection of $\Gamma$ to the plane of the picture is essentially the quiver diagram of the theory, as we will now explain.

The nodes of the quiver correspond to the intersection points of the edges of $\Gamma$.
Consider a path between two edges of  $\Gamma$ in ${\mathbb R}^3$, together with a $T^2$ fiber over it. This  gives a closed three cycle in the total space.\footnote{If the two lines meet in the base space, the three-cycle obtained in this way
can be shrunken to a point. If they don't, it generates a homology class in $H_3(X,{\bf Z})$.} The three cycles that minimize the volume come from paths of minimal length; they map to intersection points in the planar projection.
Three-cycles that arize in this way are either $S^3$'s or ${\mathbb Z}_n$ orbifolds thereof. If ${\vec v}_{L,R}$ are the vectors corresponding to 2 intersecting edges, the order of the orbifold group is $n = | {\vec v}_L \times {\vec v}_R|$. We will wrap some number $N_{\alpha}$  of branes on each minimal three cycle, corresponding to node labeled by $\alpha$. In the refined topological string, we need to choose these to be either the $q$-branes or the ${\bar t}$-branes, so we get either the
$SU(N_{\alpha})_q$ or $SU(N_{\alpha})_{\bar t}$ refined Chern-Simons theory when we compute the index. As we showed in section 3, at large $N$ the difference between the $q$-branes and ${\bar t}$-branes on cycles that undergo transitions vanishes. Thus, we may as well take all the branes to be $q$-branes, to keep things simple.

For every edge of $\Gamma$ between the nodes we get bifundamental matter. This comes about as follows.
In the Calabi-Yau geometry, these intervals along the edges lift to holomorphic annuli.  The $S^1$ boundaries of M2 branes wrapping the annuli are charged under the Chern-Simons gauge groups, and lead to matter supported on knots in the three manifolds. In general, the knots are linked, where the linking is determined by the Calabi-Yau.\footnote{Note that ordinarily, the quiver Chern-Simons theory would not be topological, as coupling to matter would require a metric. Here, the metric is not needed, precisely because the matter is localized on knots. For a pair of bifundamentals $Q, \tilde Q$, the coupling to the gauge fields in $\oint_K {\rm Tr} \tilde{Q} d_A Q$.}
 The effect of integrating out the bifundamental matter is captured by the refined Ooguri-Vafa operators from sections 2 and 3. Which operator we get depends on both the geometry of the Calabi-Yau, and the branes the annuli end on. Here, all the branes wrap compact cycles, so in place of ${\cal O}_{qq}$ operators, one has to use ${\cal O}^*_{qq}$, as we explained in section 2. This generalizes the usual notion of the quiver, the {\it nodes have topology} associated to them; the partition function will depend on the topological type of the three manifold at each node. If the node corresponds to a lens space other than an $S^3$, we also need to choose a flat connection that will break the gauge group to a subgroup. In the next section, we will do one instructive example, in detail.

\section{ An Example: A Calabi-Yau Containing Local ${\mathbb P}^2$}

\begin{figure}
  \begin{center}
    \includegraphics[width=0.5\textwidth]{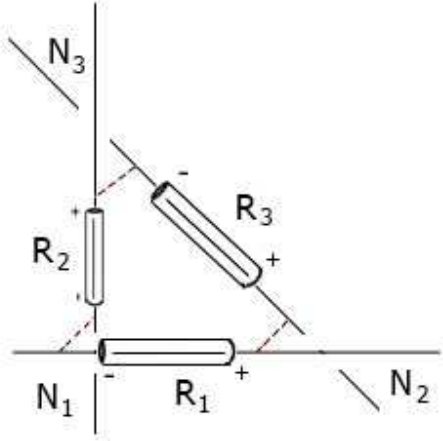}
  \end{center}
    \caption{The graph $\Gamma$, with holomorphic annuli around edges schematically shown.}\label{BeforeTrans}
\end{figure}

Consider the graph $\Gamma$ with three components, three edges are labeled as in the figure (\ref{BeforeTrans}), 

$$ v_1= (1,0), \qquad v_2= (0,1), \qquad v_3 = (-1,1).$$
 The edges intersect pairwise and give there minimal cycles. Since $|v_i \wedge v_j | = 1$, all the cycles are topologically $S^3$'s.
Now introduce $N_{1}$, $N_{2}$, $N_{3}$ $q$-branes on the three $S^3$'s.

The M2 branes wrapping holomorphic annuli give bifundamental matter. To fully specify the theory, we need to choose the orientations of the propagators; this choice is related to choosing the charges of the M2 branes under the R-symmetry.
From the figure, the propagators along the first and the third edges are
$$
{\cal O}_{q{\bar q}}(V_i, V_j;\Lambda) = \sum_R {\Lambda^{|R|}} \;iM_R(V_i)/g_R\;M_R(V_j^{-1})
$$
Here, the boundary labeled by $+$ is weighted by $V$, and the boundary labeled by $-$, by $V^{-1}$. This corresponds to the fact that the $+/-$ label two different orientations of the $S^1$ boundary of the annulus; the positive and negative oriented boundaries couple to the gauge field a on the three manifold differently. For the purposes of evaluating the expectation values, we need to expand everything in Macdonald functions. So, let us define a matrix $iG^{RQ}$, such that:
$$
{\cal O}_{q{\bar q}}(V_i, V_j;\Lambda) = \sum_{RQ} {iG}(\Lambda)^{QR} M_Q(V_i) M_R(V_j^{-1})
$$
It is easy to find an explicit expression for $iG(\Lambda)^{RQ}$; it is a symmetric, but not a diagonal matrix.

The propagator on the edge $2$ is
$$
{\cal O}^*_{qq}(V_2, V_1; \Lambda) = \sum_R {\Lambda^{|R|}} \;M_R(V_2)\;M_R(V_1^{-1})/G_R.
$$
To unify two notations in the two case, lets denote this by

$$
{\cal O}^*_{qq}(V_2, V_1;\Lambda) = \sum_R G(\Lambda)^{RQ}\;M_R(V_2)\;M_R(V_1^{-1}).
$$
where

$$G(\Lambda)^{RQ} = {\Lambda^{|Q|}/ G_Q} \;{\delta^R}_Q.
$$

The three manifolds corresponding to the nodes of the quiver arise from degenerations of the $T^2$ fiber over the edges. We can think of them as glued together from solid tori with an $SL(2,{\mathbb Z})$ transformation of their boundaries. The  $SL(2,{\mathbb Z})$ transformation $K_{ji}$ tells us how to relate the vanishing one-cycle$v_i$ and the finite cycle $f_i$  over the $i$'th edge, to that of the $j$-th edge. Let $K_i = (v_i, f_i)$, then

$$
K_{ji} = K_j^{-1} K_i.
$$
does the job: it manifestly takes a pair $K_i = (v_i, f_i)$ of the edge and framing vectors, from that of $i$'th edge to $j$'th edge.
In the present case, if we choose the framing vector for edges one and three to be $f_{1,3} = (0,\pm 1)$, and for edge two $f_2=(-1,0)$, then

$$K_1 = \left(\begin{array}{cc}1 &0\\
0 &\;1\end{array}\right), \qquad K_2 = \left(\begin{array}{cc}0 & -1\\
1&0\end{array}\right), \qquad K_3 = \left(\begin{array}{cc}1 &0\\
-1 &\;-1\end{array}\right).$$
and the gluing matrices are

$$
K_{31} = T S^{-1} T, \qquad K_{21} =S^{-1}, \qquad K_{32} = S^{-1}T^{-1}
$$

Finally, we have to compute the expectation values. Recall that insertion of $M_{R}(U)$ in a solid torus creates a state $|R\rangle$. Correspondingly, $M_{R}(U^{-1}) = M_{{\bar R}}(U)$ creates $|{\bar R}\rangle$. Putting all of this together, the refined topological string amplitude we get is

\begin{align}
 \sum_{R_i, R_i'}  {iG}^{R_1R_1'}(\Lambda_1) \; {iG}^{R_3R_3'}(\Lambda_3)\; G^{R_2R_2'}(\Lambda_2) \;
\langle {R}_3|TS^{-1}T&|R_1\rangle_{SU(N_2)_q}\;
\langle {\overline R}_2'|S|{\overline R}_1\rangle_{SU(N_1)_q}\cr
\times &\langle{\overline R}_3 |S^{-1} T^{-1}|R_2\rangle_{SU(N_3)_q}
\end{align}

One should recall that the first, second and third expectation values are computed in different gauge theories,
${SU(N_2)_q}$, ${SU(N_1)_q}$, ${SU(N_3)_q},$ respectively. The three parameters $\Lambda_1$, $\Lambda_2$, $\Lambda_3$ are all classically the same, since there is only one class in ${\mathbb P}^2$,
$\Lambda_i \sim \Lambda$. However, there can be small quantum corrections to their sizes due to the back-reaction of the branes on the $S^3$'s on the geometry. These are quantum corrections in the sense that they go away by setting $q,t=1$. Generally, for the geometry to correspond to ${\mathbb P}^2$ after the transition, with a single Kahler parameter $\Lambda$, the sizes before the transition cannot be quite the same. This phenomenon was noted in \cite{AMV} already in the unrefined case. We can simplify this further. Recall that the matrix elements are defined by

$$\langle {\overline R}_i| K|R_j\rangle = K_{{ R_i}R_j}
$$
Moreover in the refined $SU(N)$ Chern-Simons theory, charge conjugation

 $$|{\bar R}\rangle=C|R\rangle,$$
is implemented by a matrix $C=S^2$ which commutes with everything and satisfies $CS = S^{-1}$. Finally,

$$
T^{-1} = {\overline T}, \qquad S^{-1} = {\overline S}
$$
where the bar denotes an operation that sends $(q,t)\rightarrow (q^{-1}, t^{-1})$.
From this it follows that

$$
\langle {R}_3|T S^{-1}T|R_1\rangle_{SU(N_2)_q}\; = (TST)^{N_2}_{R_3R_1},
$$

$$
 \qquad\;\langle {\overline R}_2|S|{\overline R}_1\rangle_{SU(N_1)_q}=({\overline S})^{N_1}_{R_1R_2}
$$

$$
\langle {\overline R}_3 |S^{-1} T^{-1}|R_2\rangle_{SU(N_3)_q}  =({\overline{S T}})^{N_3}_{R_3 R_2}
$$

This gives our final answer for the amplitude, before the transition:

\begin{equation}
\label{eq-three}
Z_{N_1,N_2,N_3}(\Lambda)= \sum_{R_i, R_i'}  {iG}^{R_1R_1'}(\Lambda_1) \; {iG}^{R_3R_3'}(\Lambda_3)\; G^{R_2R_2'}(\Lambda_2) \;
(TST)^{N_2}_{R_3R_1}\;({\overline S})^{N_1}_{R_1R_2}\;({\overline{TS}})^{N_3}_{R_2 R_3}
\end{equation}
\subsection{Geometric Transition}
After the transition the $S^3$'s disappear, and also the branes on them. Instead, they are replaced by three ${\mathbb P}^1$'s.
The sizes of the three ${\mathbb P}^1$'s are determined by the number of the branes on the $S^3$ that gave rise to it -- the size of ${\mathbb P}^1$ is the t'Hooft coupling of the corresponding gauge group. We will see that the corresponding Kahler moduli are

$$
Q_1 = t^{N_1} \sqrt{t/q}, \qquad Q_2 = t^{-N_2} \sqrt{q/t}, \qquad Q_3 = t^{N_3} \sqrt{q/t}
$$
Moreover, we get a holomorphic four-cycle, the ${\mathbb P}^2$.
The way it arises is as follows \cite{StromingerMorrison}. Before the transition, three was a four-chain with boundaries on the three $S^3$'s.
This came about because the second $S^3$ was homologous to the sum of the third and the first $S^3$. After the transition, the $S^3$'s disappear and the four-chain's boundaries close, to give the ${\mathbb P}^2$. The Kahler parameter of the ${\mathbb P}^2$ is related to ${\Lambda_i}$ by

$$
\Lambda_1= \Lambda, \ \Lambda_2 = \Lambda, \ \Lambda_3 = \Lambda \,q/t
$$
\begin{figure}
  \begin{center}
    \includegraphics[width=\textwidth]{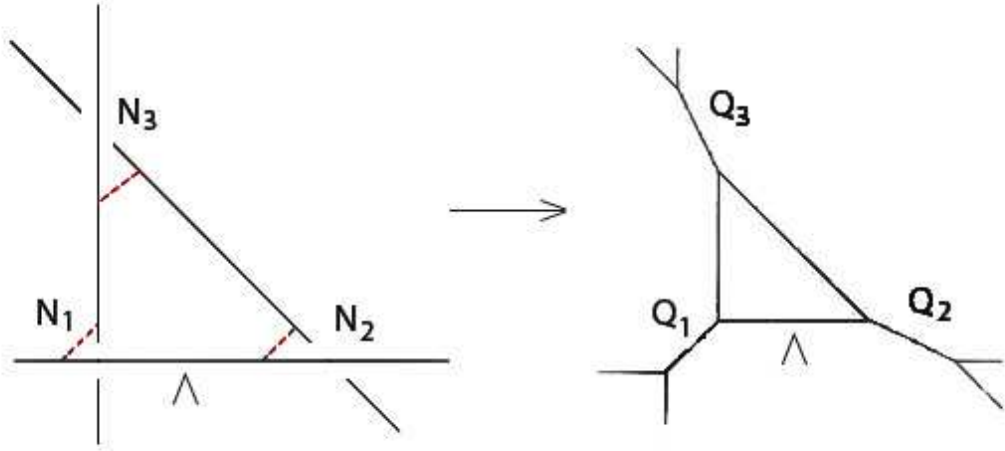}
  \end{center}
    \caption{The geometric transition results in the Calabi-Yau $X$ on the right, containing the local ${\mathbb P}^2$.\label{GeomTrans}}
\end{figure}

At large $N$, we can rewrite the partition function of the quiver Chern-Simons theory \eqref{eq-three} before the transition in terms of $Q_i$ and $\Lambda$,
\beq\label{prediction}
Z_{N_1,N_2,N_3}(\Lambda)=Z_{X}(\Lambda, Q_1,Q_2,Q_3)
\eeq
The claim of the large $N$ duality is that $Z_{X}$ is the partition function of the refined topological string on the Calabi-Yau $X$ in figure (\ref{GeomTrans}) after the transition.

To test this claim, recall that the partition function of the refined topological string has a very strict integrality property. Namely, for any Calabi-Yau where refined topological string can be defined, we expect the refined partition function to take the form \cite{GV, Vafa-Iqbal}
$$
Z ({\vec Q})= \exp( \sum_{n, {\vec d}, j_L, j_R} {(-1)^{2j_L+2j_R}{\cal N}^{\vec d}_{j_L, j_R}{\vec Q}^{n {\vec d}}\over n(q^{n/2} - q^{-n/2})(t^{n/2} - t^{-n/2})}\sum_{m_L=-j_L}^{j_L}(qt)^{nm_L}\sum_{m_R=-j_R}^{j_R}(q/t)^{m_R}
$$
where
$$
{\cal N}^{\vec d}_{j_L, j_R}
$$
are integers, counting the number of M2 branes of spin $(j_L, j_R)$ under the $SU(2)_L\times SU(2)_R$ little group of a massive particle in five dimensions. More precisely, the partition function being a twisted index, the $SU(2)_R$ is really the diagonal subgroup of the Lorentz $SU(2)_R$ times the $SU(2)_r$ R-symmetry group \cite{NO}.

We will extract, from the large $N$ expansion of  \eqref{prediction}  the BPS expansion of $Z_{X}(\Lambda, Q_1,Q_2,Q_3)$. This can be done by using for the l.h.s. of \eqref{prediction} the explicit formula \eqref{eq-three}. We will then show, that in those cases where we can do the computation using the direct approach to Gopakumar-Vafa degeneracies of \cite{GVI,GVII,KatzMayrVafa}, the numbers predicted from large $N$ duality agree with the expected ones. The results below are presented in a notation
$$
\bigoplus\limits_{j_L,j_R} {\cal N}^{\vec d}_{j_L, j_R} \big( j_L, j_R \big)
$$
i.e., every summand of the form $n (j_L,j_R)$ means that a $(j_L,j_R)$ multiplet is present in the expansion, with multiplicity $n$. We obtain, in order $\Lambda$:
\begin{center}
$(0,1)
$
\end{center}
in order $\Lambda^2$:

\begin{center}
$(0,\frac{5}{2})
$
\end{center}
in order $\Lambda^3$:

\begin{center}
$(0,3) \oplus (\frac{1}{2},\frac{9}{2})
$
\end{center}
in order $\Lambda^4$:

\begin{center}
$(0,\frac{5}{2}) \oplus (0,\frac{9}{2}) \oplus (0,\frac{13}{2}) \oplus (\frac{1}{2},4) \oplus (\frac{1}{2},5) \oplus (\frac{1}{2},6) \oplus (1,\frac{11}{2}) \oplus (\frac{3}{2},7)
$
\end{center}
in order $\Lambda^5$:
\begin{center}
$
(0,1) \oplus (0,3) \oplus (0,4) \oplus 2 (0,5) \oplus 2 (0,6) \oplus 2 (0,7) \oplus (0,8) \oplus (\frac{1}{2},\frac{5}{2}) \oplus (\frac{1}{2},\frac{7}{2}) \oplus 2 (\frac{1}{2},\frac{9}{2}) \oplus 2 (\frac{1}{2},\frac{11}{2}) \oplus 3 (\frac{1}{2},\frac{13}{2}) \oplus 2 (\frac{1}{2},\frac{15}{2}) \oplus (\frac{1}{2},\frac{17}{2}) \oplus (1,4) \oplus (1,5) \oplus 2 (1,6) \oplus 2 (1,7) \oplus 2 (1,8) \oplus (1,9) \oplus (\frac{3}{2},\frac{11}{2}) \oplus (\frac{3}{2},\frac{13}{2}) \oplus 2 (\frac{3}{2},\frac{15}{2}) \oplus (\frac{3}{2},\frac{17}{2}) \oplus (\frac{3}{2},\frac{19}{2}) \oplus (2,7) \oplus (2,8) \oplus (2,9) \oplus (\frac{5}{2},\frac{17}{2}) \oplus (3,10)
$
\end{center}
The degeneracies, up to the order $\Lambda^5$ were computed earlier, in \cite{CIV}, by a different method based on a flop of  $F_1$ and Nekrasov partition functions \cite{N, NO}. Here we present the next, $\Lambda^6$ order:

\begin{center}
$6 (0,\frac{9}{2}) \oplus 8 (0,\frac{13}{2}) \oplus 5 (\frac{1}{2},4) \oplus 6 (\frac{1}{2},5) \oplus 9 (\frac{1}{2},6) \oplus 7 (1,\frac{11}{2}) \oplus 7 (\frac{3}{2},7) \oplus 3 (0,\frac{5}{2}) \oplus (0,\frac{1}{2}) \oplus (0,\frac{3}{2}) \oplus 2 (0,\frac{7}{2}) \oplus 4 (0,\frac{11}{2}) \oplus 5 (0,\frac{15}{2}) \oplus 7 (0,\frac{17}{2}) \oplus 2 (0,\frac{19}{2}) \oplus 2 (0,\frac{21}{2}) \oplus (\frac{1}{2},1) \oplus 2 (\frac{1}{2},2) \oplus 3 (\frac{1}{2},3) \oplus 9 (\frac{1}{2},7) \oplus 10 (\frac{1}{2},8) \oplus 7 (\frac{1}{2},9) \oplus 5 (\frac{1}{2},10) \oplus (\frac{1}{2},11) \oplus (\frac{1}{2},12) \oplus (1,\frac{3}{2}) \oplus (1,\frac{5}{2}) \oplus 3 (1,\frac{7}{2}) \oplus 3 (1,\frac{9}{2}) \oplus 7 (1,\frac{13}{2}) \oplus 11 (1,\frac{15}{2}) \oplus 9 (1,\frac{17}{2}) \oplus 9 (1,\frac{19}{2}) \oplus 4 (1,\frac{21}{2}) \oplus 2 (1,\frac{23}{2}) \oplus (\frac{3}{2},3) \oplus (\frac{3}{2},4) \oplus 3 (\frac{3}{2},5) \oplus 4 (\frac{3}{2},6) \oplus 7 (\frac{3}{2},8) \oplus 10 (\frac{3}{2},9) \oplus 6 (\frac{3}{2},10) \oplus 4 (\frac{3}{2},11) \oplus (2,\frac{9}{2}) \oplus (2,\frac{11}{2}) \oplus 3 (2,\frac{13}{2}) \oplus 4 (2,\frac{15}{2}) \oplus 7 (2,\frac{17}{2}) \oplus 6 (2,\frac{19}{2}) \oplus 6 (2,\frac{21}{2}) \oplus 2 (2,\frac{23}{2}) \oplus (2,\frac{25}{2}) \oplus (\frac{5}{2},6) \oplus (\frac{5}{2},7) \oplus 3 (\frac{5}{2},8) \oplus 3 (\frac{5}{2},9) \oplus 5 (\frac{5}{2},10) \oplus 3 (\frac{5}{2},11) \oplus 2 (\frac{5}{2},12) \oplus (3,\frac{15}{2}) \oplus (3,\frac{17}{2}) \oplus 3 (3,\frac{19}{2}) \oplus 3 (3,\frac{21}{2}) \oplus 3 (3,\frac{23}{2}) \oplus (3,\frac{25}{2}) \oplus (\frac{7}{2},9) \oplus (\frac{7}{2},10) \oplus 2 (\frac{7}{2},11) \oplus (\frac{7}{2},12) \oplus (\frac{7}{2},13) \oplus (4,\frac{21}{2}) \oplus (4,\frac{23}{2}) \oplus (4,\frac{25}{2}) \oplus (\frac{9}{2},12) \oplus (5,\frac{27}{2})
$
\end{center}
For the higher orders that involve $Q_1,Q_2,Q_3$, it is convenient to present the results in the form of the following table:

\[
\begin{array}{c|llllll}
{\rm order} & \mbox{BPS spectrum} & \rule{0pt}{3mm}  \\
\hline \Lambda Q_1 & (0,\frac{1}{2}) & \rule{0pt}{5mm}  \\
\hline \Lambda^2 Q_1 & (0,2) & \rule{0pt}{5mm}  \\
\hline \Lambda^3 Q_1 & (0,\frac{5}{2}) \oplus (0,\frac{7}{2}) \oplus (\frac{1}{2},4) & \rule{0pt}{5mm}  \\
\hline \Lambda^4 Q_1 & (0,2)\oplus (0,3)\oplus 2 (0,4)\oplus (0,5) \oplus (0,6)\oplus (1,5) & \rule{0pt}{5mm} \\
 & \oplus (1,6)\oplus (\frac{1}{2},\frac{7}{2})\oplus 2  (\frac{1}{2},\frac{9}{2})\oplus2  (\frac{1}{2},\frac{11}{2})\oplus (\frac{3}{2},\frac{13}{2}) & \rule{0pt}{5mm}  \\
\hline \Lambda^3 Q_1^2 & (0,3) & \rule{0pt}{5mm}  \\
\hline \Lambda^4 Q_1^2 & (0,\frac{5}{2}) \oplus  (0,\frac{7}{2}) \oplus 2  (0,\frac{9}{2}) \oplus  (1,\frac{11}{2}) \oplus  (\frac{1}{2},4) \oplus  (\frac{1}{2},5) & \rule{0pt}{5mm}  \\
\hline \Lambda Q_1 Q_2 & (0,0) & \rule{0pt}{5mm}  \\
\hline \Lambda^2 Q_1 Q_2 & (0,\frac{3}{2}) & \rule{0pt}{5mm}  \\
\hline \Lambda^3 Q_1 Q_2 & (0,2) \oplus 2 (0,3) \oplus (\frac{1}{2}, \frac{7}{2}) & \rule{0pt}{5mm}  \\
\hline \Lambda^4 Q_1 Q_2 & (0,\frac{3}{2}) \oplus 2  (0,\frac{5}{2}) \oplus 4  (0,\frac{7}{2}) \oplus 3  (0,\frac{9}{2}) \oplus  (0,\frac{11}{2}) & \rule{0pt}{5mm}  \\
 &
\oplus  (1,\frac{9}{2}) \oplus 2  (1,\frac{11}{2}) \oplus  (\frac{1}{2},3) \oplus 3  (\frac{1}{2},4) \oplus 4  (\frac{1}{2},5) \oplus  (\frac{3}{2},6) & \rule{0pt}{5mm}  \\
\hline \Lambda^4 Q_1^3 & (0,4) & \rule{0pt}{5mm}  \\
\hline \Lambda^3 Q_1^2 Q_2 & (0,\frac{5}{2}) & \rule{0pt}{5mm}  \\
\hline \Lambda^4 Q_1^2 Q_2 & (0,2) \oplus 2  (0,3) \oplus 3  (0,4) \oplus  (1,5) \oplus  (\frac{1}{2},\frac{7}{2}) \oplus 2  (\frac{1}{2},\frac{9}{2}) & \rule{0pt}{5mm}  \\
\hline \Lambda^2 Q_1 Q_2 Q_3 & (0,1) & \rule{0pt}{5mm}  \\
\hline \Lambda^3 Q_1 Q_2 Q_3 & (0,\frac{3}{2}) \oplus 3 (0,\frac{5}{2}) \oplus (\frac{1}{2},3) & \rule{0pt}{5mm}  \\
\hline \Lambda^4 Q_1 Q_2 Q_3 &  (0,1) \oplus 3  (0,2) \oplus 7  (0,3) \oplus 7  (0,4) \oplus  (0,5) \oplus  (1,4) & \rule{0pt}{5mm}  \\
 &
\oplus 3  (1,5) \oplus  (\frac{1}{2},\frac{5}{2}) \oplus 4  (\frac{1}{2},\frac{7}{2}) \oplus 7  (\frac{1}{2},\frac{9}{2}) \oplus  (\frac{3}{2},\frac{11}{2})
 & \rule{0pt}{5mm}  \\
\end{array}
\]
and so on. Provided enough computer time and memory, our explicit formulas allow to compute the BPS spectra up to any desired order.

There is a direct way to compute the degeneracies of BPS states, as explained by Gopakumar and Vafa in \cite{GVI, GVII} and further developed in \cite{KatzMayrVafa} (see also \cite{Mirrorbook} for a pedagogical introduction). This direct way of computation is based upon a fact that BPS states in order $\Lambda^d$ are in one-to-one correspondence with homology classes of the moduli space of degree $d$ curves equipped with a $U(1)$ bundle. The computation of these homologies is not always easy; however, some cases are elementary. In particular, it is easy to compute the leading multiplet in the expansion, that is, the multiplet with the highest $j_L$ spin. It is known \cite{Mirrorbook} that the highest $j_L$ spin in order $\Lambda^d$ is
$$
j_L^{max} = g/2
$$
where $g = \dfrac{(d-1)(d-2)}{2}$ is the genus of a generic degree $d$ curve. Moreover, the BPS states with such highest spin have a simple interpretation as cohomologies of the moduli space of degree $d$ curves in ${\mathbb P}^2$, with no bundles involved. This moduli space is elementary to compute: a degree $d$ has plane curve has a form
$$
\sum\limits_{i+j+k=d} c_{ijk} x_1^i x_2^j x_3^k = 0
$$
and has $(d+1)(d+2)/2$ coefficients defined up to an overall rescaling; the moduli space of such curves is therefore ${\mathbb C}{\mathbb P}^{d(d+3)}$, whose total cohomology (the sum of all Betti numbers) is $d(d+3) + 1$. To agree with this prediction, the highest $j_L$ multiplet has to have a form
$$
\mbox{order} \ \Lambda^d: \ \ \ \left( \dfrac{(d-1)(d-2)}{4}, \dfrac{d(d+3)}{4} \right)
$$
One can see, that this agrees with our results, for $d = 1,2,3,4,5$. To compute the subleading contributions with lower $j_L$, one would have to deal with more complicated moduli spaces -- see \cite{Mirrorbook} for some examples.

Including into consideration the non-zero degrees in $Q_1,Q_2,Q_3$ is also easy, up to degree one in each $Q_i$. Consider, say, order $\Lambda^d Q_1$. In this case, the BPS states with highest $j_L$ spin correspond to homologies of the moduli space of degree $d$ plane curves in ${\mathbb P}^2$ that pass through a given point -- this is the point where the ${\mathbb P}^1$ of class $Q_1$ attaches. Such curves have $(d+1)(d+2)/2$ coefficients defined up to an overall rescaling and subject to one linear constraint; topologically, this is just ${\mathbb C}{\mathbb P}^{d(d+3)-1}$, whose total De Rham cohomology is $d(d+3)$. We then obtain the highest $j_L$ multiplet
$$
\mbox{order} \ \Lambda^d Q_1: \ \ \ \left( \dfrac{(d-1)(d-2)}{4}, \dfrac{d^2 + 3d - 2}{4} \right)
$$
One can see, that this agrees with our results, for $d = 1,2,3,4$. Similarly, in the cases $\Lambda^d Q_1 Q_2$ and $\Lambda^d Q_1 Q_2 Q_3$ we have to consider the moduli space of degree $d$ plane curves that pass through a pair and a triple of distinct points, respectively; the moduli spaces in these two cases are ${\mathbb C}{\mathbb P}^{d(d+3)-2}$ and ${\mathbb C}{\mathbb P}^{d(d+3)-3}$. To match these algebraic geometry considerations, the highest $j_L$ multiplets have to have a form
$$
\mbox{order} \ \Lambda^d Q_1 Q_2: \ \ \ \left( \dfrac{(d-1)(d-2)}{4}, \dfrac{d^2 + 3d - 4}{4} \right)
$$
$$
\mbox{order} \ \Lambda^d Q_1 Q_2 Q_3: \ \ \ \left( \dfrac{(d-1)(d-2)}{4}, \dfrac{d^2 + 3d - 6}{4} \right)
$$
Again, this agrees with our results above, for $d \leq 4$.

\subsection{Cutting the Semi-Local ${\mathbb P}^2$ Amplitude into Vertices}
It is an interesting question whether the amplitude in \eqref{eq-three}  can be written in terms of topological vertex amplitudes of section 4.
We will now show that this is indeed the case. In the next section, we will return to this question in a more general setting, and provide an interpretation of
this result in terms of a new refined topological vertex formalism.

Starting with the amplitude \eqref{eq-three} before the transition, we can express it in terms of refined topological vertices using the following identities. We have:

$$
(S)_{A,B}= \sum\limits_{RR'} g^{RR'} \; {\cal C}_{R, A, B}\; {\overline {\cal C}}_{R, 0, 0} \;\big( Q v \big)^{|R|} .
$$
On the left hand side is the $S$ matrix of $SU(N)_q$ refined Chern-Simons theory. We have expressed it through gluing the refined topological vertices, where $Q= t^{-N} v$.  On the right hand side, the topological vertices are glued with a diagonal metric
$$g^{R R'} = g_R {\delta^{R}}_{R'}$$
is the infinite $N$ limit of the Macdonald metric $G^{RR'}.$ It is diagonal with eigenvalue that is usually denoted by $g_R$. This is an algebraic identity, a version of which is proven in \cite{Iqbal}.

The complex conjugate of this relation, corresponding to replacing $(q,t)$ with $(q^{-1}, t^{-1})$ is
$$
({\overline S})_{A,B} =
\sum\limits_{R,R'} g^{RR'} \; {\overline{\cal C}}_{R, A, B} \;{\cal  C}_{R', 0, 0} \;\big( Q v \big)^{|R|}
$$
Here $Q$ is the complex conjugate of the above, $Q = t^{N} v^{-1}$.  Note that $g_R$ is real. In our theory, we will need the large $N$ expansions of both $S$ and $\overline{S}$. Naively, for any fixed value of $t$ only one or the other expansion would be valid. However, using the $q/t$ exchange symmetry of the refined Chern-Simons theory at large $N$, we could trade the offending $q$-branes for $t-$branes. Then, since $q$ and $t$ are independent, we would always be able to make both expansions convergent.

Finally, we have
$$
\sum_{A' B'} {iG}^{AA'}({\overline S})_{A'B'} G^{BB'}=
\sum\limits_{\stackrel{A,R,B}{A'B'Y'}}g^{AA'} g^{RR'} g^{BB'} \; {\cal C}_{A', R, B'}\;  {\overline{\cal  C}}_{0, R, 0} \big( Q v \big)^{|R|} f^{-1}_R
$$
where $Q = t^{N} v$. Here $f_R$ is the infinite $N$ limit of the framing matrix,
$$
f_R = (-1)^{|R|} \ q^{||R||/2} \ t^{-||R^T||/2}
$$
Note that the left hand side involves the Macdonald metric $G_{RR'}$ and the right hand side only the infinite $N$ variant, $g_{RR'}$.

Using these relations, the large $N$ limit of the amplitude in \eqref{eq-three} can be rewritten directly in terms of refined topological vertices, as follows
\begin{align}
Z_{{\mathbb P}^2}\big( \Lambda, Q_1, Q_2, Q_3 \big) = \sum\limits_{\it{indices}} \it{vertices} \times \it{gluing\; factors} \times \it{Kahler \;parameters}
\label{ZP2vertex}
\end{align}
where
$$
{\it indices} = R_{i}, R'_{i}, Y_i , Y_i', \qquad i=1,2,3
$$
$$
{ \it vertices} = {\overline {\cal C}}_{Y_1, R_1, R_2^{\prime}} \;\;{\cal C}_{Y_2, R_3, R_1^{\prime}} \;\;{\cal C}_{R_3^{\prime}, Y_3, R_2}\;\;{\cal  C}_{Y_1^{\prime}, 0, 0}\; \;{\overline {\cal C}}_{Y_2^{\prime}, 0, 0} \;\;{\overline {\cal C}}_{0, Y_3^{\prime}, 0}
$$
$$
{\it gluing \; factors} =
\big(i g \; f\big)^{R_1, R_1^{\prime}}
\big( g\; f \big)^{R_3, R_3^{\prime}}\;
\big(g {f}^{-2}\big)^{R_2, R_2^{\prime}}\;
\big(g \big)^{Y_1, Y_1^{\prime}}\;
\big(g\big)^{Y_2, Y_2^{\prime}}\;
\big(g\big)^{Y_3, Y_3^{\prime}}
$$
$$
\it{Kahler \;parameters} =
\big(\Lambda\big)^{|R_1|} \ \big(\Lambda v^2\big)^{|R_3|} \ \big(\Lambda v\big)^{|R_2|} \ \big( Q_1 v \big)^{|Y_1|} \ \big( Q_2 v \big)^{|Y_2|} \ \big( Q_3 v \big)^{|Y_3|}
$$
In the above formulas,
$$g^{R R'} = g_R {\delta^{R}}_{R'}$$
and
$$(i g)^{R R'}$$
are the infinite $N$ limits of  $G^{RR'}$ and $(iG)^{RR'}$ (the later is independent of $N$ in fact).

It is easy to show that, setting $q=t$, this reduces to the ordinary topological vertex formalism for this geometry, after we adjust the framing and orientation of the vertex factors, so as to restore the cyclic symmetry ${\mathbb Z}_3$ symmetry. For $q\neq t$, however, the vertex formulation of this geometry is genuinely new.

\section{Towards the Topological Vertex Formalism}

Ordinary topological vertex led to a very elegant way of computing topological string amplitudes on arbitrary toric Calabi-Yau manifolds. By cutting a Calabi-Yau into ${\mathbb C}^3$ pieces, one is able to obtain the topological string partition function by gluing the ${\mathbb C}^3$ amplitudes. The building block of the theory is the topological vertex, i.e.  the topological string partition function on ${\mathbb C}^3$ with three stacks of Lagrangian branes. The fact the topological vertex indeed computes the topological string amplitudes was proven recently in \cite{OkounkovOblomkovMaulikPandharipande}.

In the refined topological string case, a topological vertex formalism was developed in \cite{AK, CIV}\footnote{See \cite{AwataFeigin} for recent nice work generalizing \cite{IntegrableHierarchies}.} for those toric geometries that lead to  ${\cal N}=2$ $SU(N)$ (quiver) gauge theories in four dimensions. The geometries have the property that one can choose a "preferred direction" in the toric graph $\Gamma$ such that at each vertex, one of the three legs points along this direction. Naturally, one would like to understand whether there is a refined vertex formalism that will allow one to compute the refined topological string theory amplitudes on arbitrary toric Calabi Yau manifolds.

In the previous section, using large $N$ transitions, we computed the refined topological string partition function on a local Calabi-Yau manifold $X$, which does not fall into this class. We moreover showed that the resulting partition function decomposes in terms of the refined topological vertices.  In this section, we will give a physical interpretation to the expression in terms of a new refined vertex formalism that should allow one to compute the refined topological string amplitudes on any toric Calabi-Yau.  We will explain what we believe are some of the main features of the formalism, though we will not fill in all of its details. It is easy to show that the new vertex formalism reduces to the results of \cite{CIV, AK} for $SU(N)$ geometries, but also extends to the more general class presented in section 6. In the rest of this section, we begin by discussing the general aspects of the formalism, and then show how it reproduces the results of section 6.

Recently, in a beautiful work \cite{NO2}  a new refined topological vertex that depends on four continuous parameters (one instanton weight and three weights of the equivariant $T^3$ action on $Y$) in addition to the three Young diagrams. This vertex formalism is apriori different from that in \cite{TV}, in that it comes not from Calabi-Yau manifolds with M5 branes wrapping Lagrangian cycles, but from a 6 dimensional Donaldson-Thomas theory on $Y$, refining the approach in \cite{QF}. The two descriptions ended up being the same in the context of the ordinary topological string, but in the refined case, they need not be. There is a way to specialize the equivariant weights in \cite{NO2} a way that ends up depending on a choice of a $U(1)$ vector field acting on the Calabi-Yau. It is this description that is closest to the picture we will present below, albeit it requires infinitely many vertices. Understanding the relation of the two approaches is a very interesting question, currently under investigation \cite{ASO}.

\subsection{The vertices }

To write down the vertex amplitudes in section 4, we had to choose an orientation of  the toric legs. This choice of orientation can be traced back to the orientations of the boundaries of the annuli in figures (\ref{TopVertex}) and (\ref{TopVertexConjugate}). In the ordinary topological string,  the choice of the orientation affects the amplitudes in a very simple way. Changing the orientation of a leg is simply a transposition of the corresponding representation, making it indistinguishable from flipping the Lagrangian brane to an anti-brane. It is possible to choose the orientations of the legs and framings so that the vertex amplitude has a ${\mathbb Z}_3$ symmetry that permutes the legs cyclically. Moreover, the closed string amplitudes do not depend on any such choices: the topological vertex formalism of \cite{TV}  assigned an amplitude to a graph $\Gamma$, and hence the Calabi-Yau in a unique way. The graph $\Gamma$ there consisted simply of lines of integer slope, connected with trivalent vertices.

\begin{figure}[ht]
  \begin{center}
    \mbox{\subfigure{\includegraphics[width=2in]{pic3a.eps}}\quad
\subfigure{\includegraphics[width=2in]{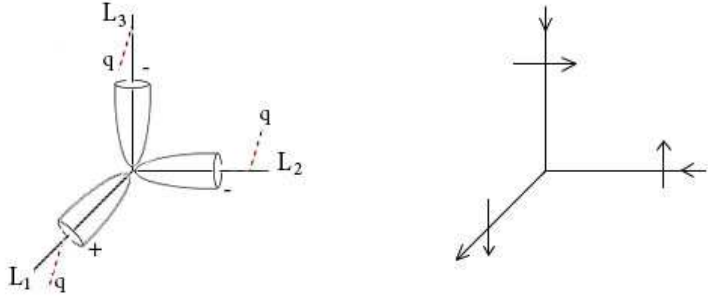} }}
  \end{center}
    \caption{The topological vertex. \label{VertexGraph}}
\end{figure}

\begin{figure}[ht]
  \begin{center}
    \mbox{\subfigure{\includegraphics[width=2in]{pic3b.eps}}\quad
\subfigure{\includegraphics[width=2in]{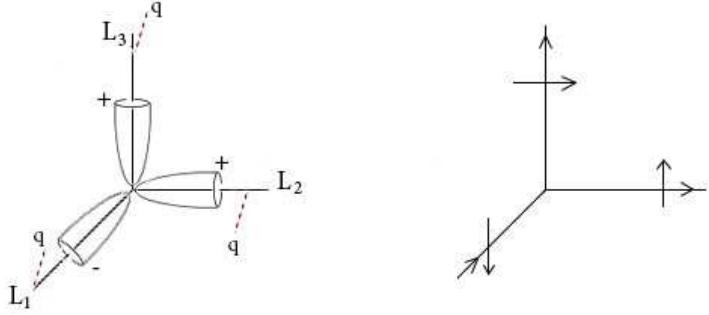} }}
  \end{center}
    \caption{The conjugate topological vertex. \label{VertexGraphConjugate}}
\end{figure}

In the refined topological string, the choice of orientations of the legs is physical, in the sense that it does not drop out of the amplitudes in the closed string in general, and changes the open string amplitudes in a rather less trivial way. Moreover, it is unrelated to the choice of branes or anti-branes.\footnote{\label{footnote} In the closed string case, the refined topological string amplitude does not depend only on the Calabi-Yau geometry any time the Calabi-Yau has holomorphic curves that can run off to infinity. In this cases, the amplitudes are not invariant under the $SU(2)_L\times SU(2)_R$ symmetry -- the BPS states do not transform in complete multiplets of $SU(2)_R$. Consequently, the choice of a $U(1)_R\subset SU(2)_R$ subgroup needed to define the theory matters. A simple example of this phenomenon is ${\cal O}(0)\oplus {\cal O}(-2)\rightarrow {\mathbb P}^1$. The moduli space of the ${\mathbb P}^1$ is non-compact in the ${\cal O}(0)$ direction.
In this case, the BPS content is that of a single state, coming from the M2 brane wrapping the ${\mathbb P}^1$ once, and transforming either as a spin $1/2$ up or spin $1/2$ down of $SU(2)_R.$ The choice of the $U(1)_R$ symmetry can trade one for the other. Were we to compactly the ${\cal O}(0)$ direction to a ${\mathbb P}^1$, we would get the complete spin $1/2$ multiplet.} Because of this, it is not possible to obtain a vertex that is completely invariant under the cyclic ${\mathbb Z}_3$ permutations of the legs. This being the case, we may as well work with a less then symmetric configuration of branes, as in the figures (\ref{VertexGraph}) and (\ref{VertexGraphConjugate}).

In section 4 we derived, from refined Chern-Simons theory two inequivalent vertex amplitudes \eqref{VAK} and  \eqref{VAKc}.
It is helpful to summarize the vertex amplitudes and the choices involved in a more practical notation.
To the vertex

$$
{ {\cal C}}_{R_1 R_2 R_3}(q,t) = \sum\limits_{R}
 g_R\;  v^{-2|R|}
\ iM_{R_1 / R}(t^{-\rho} q^{-R_3})
\ M_{R_2 / R}(t^{\rho} q^{R_3})
\ M_{R_3}(t^{-\rho}).
$$
we will associate the graph in figure (\ref{VertexGraph}).
It is important to keep track not only of the orientations of the legs, but also of the branes. We will take all the branes in this section to be $q$-branes, but we need to remember whether they are coming in or out of the plane of the diagram; in the unrefined limit, these become brane/anti-brane pairs. We will do this as follows: for any one leg with the edge vector $v_i$ (taken in the orientation of the diagram), when we draw the framing arrow $f_i$, we will chose its orientation such that it points to the right of the edge, $v_i\wedge f_1 = +1$ if the brane is coming out of the paper, and to the left of the edge, i.e.  $v_i\wedge f_1 = -1$ if it is coming into the paper.

For the second vertex
$$
{\overline {\cal C}}_{R_1 R_2 R_3}(q,t) = \sum\limits_{R}
 g_R
\ iM_{R_1 / R}(t^{\rho} q^{R_3})
\ M_{R_2 / R}(t^{-\rho} q^{-R_3})
\ M_{R_3}(t^{\rho}).\;\;\,\,\,\,\,
$$
this gives the figure (\ref{VertexGraphConjugate}).

\subsection{A Morse Flow on $\Gamma$}
It is easy to see, in examples, that the choice of orientation of a leg of a graph $\Gamma$ is related to a choice of the $U(1)_R$ symmetry action on the Calabi-Yau (see footnote \ref{footnote}). Thus we cannot expect to be able to choose the orientations of the legs at will, independently of each other. A natural way to connect a choice of a $U(1)$ action of the Calabi-Yau, with  a choice of orientation of the edges of $\Gamma$ is as follows.\footnote{This discussion is inspired by the upcoming work of A. Okounkov and N. Nekrasov \cite{NO2}.} A $U(1)_R$ symmetry is a $U(1)$ action on the Calabi-Yau that acts non-trivially on the holomorphic three-form. In the refined topological string $\Omega^{3,0}$ has to transform as $\Omega^{3,0}\rightarrow t/q \;\Omega^{3,0} $ as we go around the $S^1$. This is needed to cancel the transformations of  $z_1,z_2$ directions induced by the $\Omega$ background in \eqref{Omega}. A choice of the $U(1)$ action corresponds to a choice of a Morse flow on the Calabi-Yau, which in turn is related to a choice of a vector $\zeta$, in the plane of the toric graph. A generic choice of the Morse flow can be used to  give an orientation to the edges of $\Gamma$ to be along the flow.\footnote{  A $U(1)$ action on the toric Calabi-Yau is captured by a vector $\sum_{i} n_{\zeta, i} {\vec w}_i$ where ${\vec w}_i$ are vectors in ${\mathbb Z}^3$ associated to the coordinates $X_i$ of the toric variety. The corresponding Morse function is $\sum_i n_{\zeta,i} |X_i|^2$. This generates a $U(1)$ action that takes $X_i \rightarrow \lambda^{n_{\zeta, i}} X_i$.
For a toric Calabi-Yau, all of the vectors ${\vec w}_i$ lie in a plane, distance $1$ from the origin of ${\mathbb Z}^3$, i.e. ${\vec w}_i$  have the form $(*,*,1).$  Projecting the ${\vec w_i}$ to this plane, and choosing a triangulation, we get the toric diagram coresponding to the Calabi-Yau. The vector $\zeta$ is the projection of $\sum_{i} n_{\zeta, i} {\vec w}_i$  to this plane. The graph $\Gamma$ is the dual graph to the toric diagram. It is then easy to see why a choice of generic $\zeta$ determines an orientation to the edges of $\Gamma$.} Namely, for every edge $v_i$, we choose its orientation such that
$$
(\zeta, v_i)>0.
$$
This defines orientations of the edges for a generic choice of $\zeta$, i.e. whenever $\zeta$ is not orthogonal to any of the edges.  This suggests that the refined vertex formalism has chamber structure: in a given chamber, the orientations are independent of a specific choice of $\zeta$. The walls of the chambers are determined by choices of $\zeta=\zeta_i$ such that $(\zeta_i, v_i)=0$ for some edge $v_i$ of $\Gamma$. Crossing the wall, the edges along $v_i$ flip orientation. The choice of this vector field is also what breaks the would-be cyclic ${\mathbb Z}_3$ symmetry of the topological vertex.

\begin{figure}
  \begin{center}
    \includegraphics[width=0.75\textwidth]{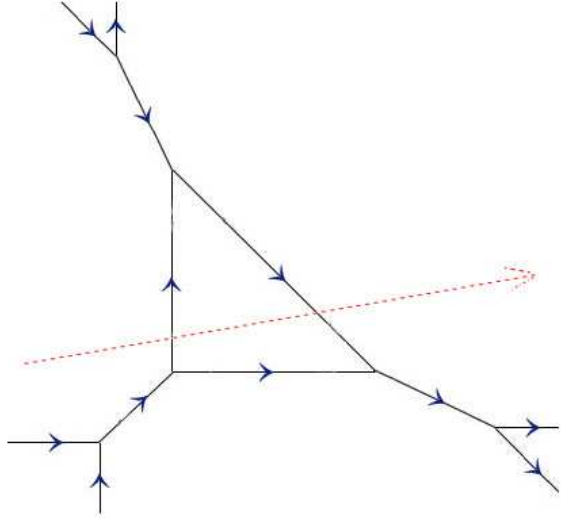}
  \end{center}\label{P2M}
    \caption{The Morse flow, and induced orientations. \label{Orientations} }
\end{figure}

As an example, consider the Calabi-Yau $X$, containing the local ${\mathbb P}^2$, which we studied in section 6. We can choose a vector field as in the figure (\ref{Orientations}), and this assigns the orientations of edges as shown.

\subsection{Sewing the Vertices}

Next, let us explain the general features of cutting and gluing of the amplitudes. The basic idea in \cite{TV} was to cut up a Calabi-Yau into ${\mathbb C}^3$ pieces by placing toric Lagrangian brane/anti-brane pairs on the legs. With branes present, one is studying maps to ${\mathbb C}^3$ with suitable boundary conditions. Canceling off the boundaries, by canceling off brane/anti-brane pairs, one can recover the amplitudes of the Calabi-Yau. In this way, by cutting and gluing, topological string amplitudes can be obtained from the ${\mathbb C}^3$ pieces -- the topological vertices. One necessary condition for the gluing is that the $S^1$ boundaries on the two pieces that we glue have to be oriented oppositely, to be able to cancel. All these considerations are purely topological, so they should naturally extend to the refined case.

\begin{figure}[ht]
  \begin{center}
\includegraphics[width=4.5in]{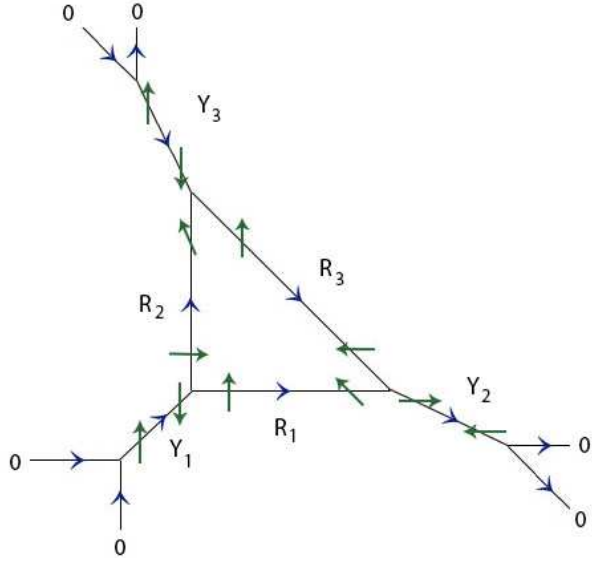}
  \end{center}
    \caption{The topological vertex formalism for the semi-local ${\mathbb P}^2$. \label{P2vertexstyle} }
\end{figure}

Like in \cite{TV}, we have kept track of the orientation of the boundary by the edge arrows. Thus, naturally, we can only glue an incoming to an outgoing edge. Applying this to the Calabi-Yau $X$, containing the ${\mathbb P}^2$, we get the following decomposition, see figure (\ref{P2vertexstyle}).

Note that by design, this divides the Calabi-Yau into one of two different kinds of vrtices, depending on whether they have one or two incoming edges. These are exactly the vertex factors of the amplitude we derived for this geometry in the previous section, using geometric transitions. In the case of $X$, this gives

$$
{ \it vertices} = {\overline {\cal C}}_{Y_1, R_1, R_2^{\prime}} \;\;{\cal C}_{Y_2, R_3, R_1^{\prime}} \;\;{\cal C}_{R_3^{\prime}, Y_3, R_2}\;\;{\cal  C}_{Y_1^{\prime}, 0, 0}\; \;{\overline {\cal C}}_{Y_2^{\prime}, 0, 0} \;\;{\overline {\cal C}}_{0, Y_3^{\prime}, 0}
$$

Next, we will explain the gluing factors.
In the refined case, we cannot use brane/anti-brane pairs for cutting and gluing, as this would break supersymmetry and supersymmetry is crucial to be able to define the theory. In their place, we want to use pairs of $q$ branes, ending on opposite sides of the leg. We can think of the branes pointing out of the page as $q$ branes, while those pointing into the page as ${\bar q}$ branes, effectively. While the branes preserve the same supersymmetry, they have opposite charge in homology -- in the sense that they can pair up to make a Lagrangian of topology $T^2\times {\mathbb R}$ that can be moved off to infinity \cite{AV12, AVK}.  In terms of the trivalent graphs of figures (\ref{VertexGraph}) and (\ref{VertexGraphConjugate}), this corresponds to gluing an incoming to an outgoing leg where the framing arrows {\it anti-align}. In this case, the gluing simply corresponds to multiplying the left with the right side and integrating. To see the effect to this,
consider, in a general setting, gluing two amplitudes $Z^q$ and ${Z'}^{q}$, where

\beq\label{lq}
Z^{q}(V)= \sum_Q {Z^{q}}_Q \; M_Q(V^{-1})/g_Q
\eeq
corresponds the geometry on the left, with $q$-branes pointing out of the page,  and

\beq\label{lbq}
{Z'}^{{ q}}(V)= \sum_Q {Z'^{q}}_{\;\;Q}\; M_Q(V)/g_Q
\eeq
comes from the geometry on the right, with $q$ branes pointing in to the page.
Gluing them together, corresponds to integrating $Z^{q}(V)\; Z^{{q}}(V)$ with Macdonald measure.\footnote{More precisely, to define the integral we first have to take a finite number $\#$ of branes, integrate, and then take $\#$ to infinity. With $\#$ branes, setting the holonomies equal, and integrating
with Macdonald measure $d_{q,t}V = d^\#V \prod_{i\neq j}\prod_{n=0}^{\infty} {1-q^n V_i/V_j\over 1-q^n t V_i/V_j}$, we get the result. This is because as $\int  d_{q,t}V M_R(V) M_Q(V^{-1}) = G_R \delta^{R}_{Q}$, where $G_R$ is the $\#$-dependent Macdonald metric. Taking $\#$ to infinity in the end, $G_R$ reduces to $g_R$, and the result follows.}
This gives $
\sum_Q {Z^{q}}_Q\;\; {{ Z'}^{{ q}}}_{\;\;Q}/{g_Q }.
$
We will write this, using $g^{QQ'}={\delta^Q}_{Q'}{/g_Q} $, as

\beq\label{qqglue}
\sum_{QQ'} {Z^{q}}_Q\; g^{QQ'}\;{{Z'}^{{q}}}_{\;\;\;Q'}
\eeq
Going back to the figure (\ref{P2vertexstyle}), it is easy to see that along all but one edge (the $R_1$ edge), the framing vectors are pointing in opposite directions -- after adjusting framing, they become anti-aligned-- and we are gluing $q$ to ${\bar q}$ branes. Thus we have explained the $g^{RR'}$ factors in the equation (\ref{ZP2vertex}).

This leaves us with explaining the edge labeled by $R_1$. On this edge, framing vectors can be made parallel, after adjusting framing.\footnote{Note that, while framing ambiguity of \cite{TV} does allow one to change the direction of the framing vector, it does not allow us to change its orientation: change of framing preserves $v_i\wedge f_i$, per definition.}  This means that we are gluing a pair of branes both of which are pointing out of the paper.
We cannot glue $q$-branes to $q$-branes directly, as there is no way to cancel them out.
Instead, we can simply introduce a propagator: in this case an annulus two stacks of ${\bar q}$ branes, i.e. $q$ branes both pointing into the page. Then, we can glue twice, in each case pairing the $q$ branes from the vertex with the ${\bar q}$ branes of the propagator.
We can think of the propagator as a patch of Calabi-Yau of the topology of  ${\mathbb C}^*\times {\mathbb C}^2$, whose toric diagram is a line.
The corresponding amplitude,  ${\cal O}_{{ q} {\bar q}}(V, {\tilde V})$ is given in (\ref{qbarq}) in section 4:

$$
{\cal O}_{{ q} {\bar q}}(V,{\tilde V}) = \sum_{R,R'} {(ig)}^{RR'} M_R(V) M_{R'}({\tilde V}^{-1})
$$
To glue an \eqref{lq} to an amplitude

\beq\label{laq}
{Z'}^{{\bar q}}(V)= \sum_Q {Z'^{\bar q}}_{\;\;Q}\; M_Q(V^{-1})/g_Q
\eeq
corresponding to ${\bar q}$-branes we multiply the three factors together
  $Z^{q}(V){\cal O}_{{q} {\bar q}}(V,{\tilde V}) { Z'}^{\bar q}({\tilde V})$, and glue by integrating over  $\int d_{q,t}V \,\int d_{q,t} {\tilde V}.$
This gives
\beq\label {qbqglue}
\sum_{QQ'} {Z^{q}}_Q\; (ig)^{QQ'}\;{{Z'}^{{\bar q}}}_{\;\;\;Q'}.
\eeq
This is exactly the propagator on leg $R_1$, up to the framing factor needed to make the branes exactly parallel.\footnote{Using the same idea, we can also glue  $q$-branes with ${\bar t}$-branes, by introducing an annulus with a $q{\bar t}$ brane pair.
$$
{\cal O}_{{ q} {\bar t}}(V,{\tilde V}) = \sum_{R} (-1)^{|R|}M_R(V) {\hat M}_{R^T}({\tilde V}^{-1}).
$$
Integrating over  $\int d_{q,t}V \,\int d_{t^{-1}, q^{-1}} {\tilde V} $
gives

\beq
\sum_{R} (-1)^{|R|}\; {Z^{q}}_{R}\;{ Z'^{t}}_{\;R^T}
\eeq
This is just the gluing of the $q$-leg and the leg in \cite{CIV}. Note that in the unrefined limit, when $q=t$ the branes become the topological anti-branes of $q$ branes. Then, this is just the gluing of \cite{TV}.} Thus, we have explained all the elements of the topological vertex computation in section 6.

\acknowledgments

We have benefited from discussions with A. Iqbal, N. Nekrasov, K. Schaeffer  and C. Vafa. We thank A. Iqbal for kindly coordinating submission of his upcoming related work with ours. We are especially grateful to A. Okounkov for sharing his unpublished results and collaboration on a related project.  We are grateful to Simons Center for Geometry and Physics and the organizers of the 2011 Summer Simons workshop in Mathematics and Physics where some of this work has been done. We also thank the organizers of the Simons Foundation Symposium on "Knot Homologies and BPS States", where parts of this work were presented. This work is supported in part by the Berkeley Center for Theoretical Physics, by the National Science Foundation (award number 0855653) and by the Institute for the Physics and Mathematics of the Universe.

\pagebreak

\appendix

\section{Definitions}

\paragraph{Macdonald polynomials} $M_Y(U)=M_Y(U; q,t)$ are defined as the unique basis of symmetric functions in the eigenvalues $U_1, \ldots, U_N$ labeled by partitions (Young diagrams) $Y = (Y_1 \geq Y_2 \geq \ldots)$, that are orthogonal with respect to the Macdonald integral scalar product
\begin{align}
\Big< f, g \Big> = \int\limits_{0}^{2\pi} d x_1 \ldots d x_N \ \Delta_{q,t} \ f\big( e^{i x_1}, \ldots, e^{i x_N} \big) g\left(e^{- i x_1}, \ldots, e^{- i x_N} \right)
\end{align}
where $\Delta_{q,t}$ is the Macdonald measure
\begin{align}
\Delta_{q,t} = \prod\limits_{m = 0}^{\beta - 1} \prod\limits_{i \neq j} \left( 1 - q^{m} e^{x_i - x_j} \right)
\end{align}
The orthogonality condition states
\begin{align}
< M_Y, M_{Y^{\prime}} > = G_Y \delta_{Y, Y^{\prime}}
\end{align}
where the quantity $G_Y$ is the quadratic norm w.r.t this product:
\begin{align}
G_Y = G_{\varnothing} \prod\limits_{(i,j) \in Y} \frac{1 - t^{Y^T_j-i} q^{Y_i-j+1} }{1 - t^{Y^T_j-i+1} q^{Y_i-j} } \ \frac{1 - (t/q) t^{N-i} q^{j} }{1 - t^{N-i} q^{j}}
\end{align}
where
$$
G_{\varnothing} = N! \prod_{m=0}^{\beta-1}
\prod_{i < j}
\dfrac{
 t^{ \frac{i-j}{2} } q^{ \frac{m}{2} } - t^{ - \frac{i-j}{2} } q^{ - \frac{m}{2} }
}
{
 t^{ \frac{i-j}{2} } q^{ - \frac{m}{2} } - t^{ - \frac{i-j}{2} } q^{ \frac{m}{2} }
}
$$
Just as any other symmetric polynomial, any Macdonald polynomial can be always expressed as a function of the power sums $p_k = {\rm Tr} U^k = U_1^k + \ldots + U_N^k$. While this is very convenient for calculational purposes, in physics-inspired contexts, such as refined Chern-Simons or topological string calculations, it is more natural to use the holonomies notation $M_R(U) = M_R(U;q,t)$.

\paragraph{Explicit expressions} for Macdonald polynomials are often convenient for reference and comparison. With our definitions, they have a form
$$
M_1(U) = s_{1}(U), \ \ \
M_2(U) = s_2(U) + \frac{q-t}{1-tq} s_{11}(U), \ \ \ M_{11}(U) = s_{11}(U)
$$
$$
M_3(U) = s_3(U) + \dfrac{(q-t)(1+q)}{1-tq^2} s_{21}(U) + \dfrac{(q-t)(q^2-t)}{(1-tq)(1-tq^2)} s_{111}(U)
$$
$$
M_{21}(U) = s_{21}(U) + \dfrac{(q-t)(1+t)}{1-t^2q} s_{111}(U), \ \ \ M_{111}(U) = s_{111}(U)
$$
where $s_R(U) = {\rm Tr}_R(U)$ are the Schur polynomials. It can be shown that Macdonald polynomials satisfy $M_Y(U; q,t)=M_Y(U; q^{-1},t^{-1}).$

\paragraph{Cauchy identity} is an important sum rule, satisfied by Macdonald polynomials:
\begin{align}
\sum\limits_{Y} \frac{1}{g_Y} M_Y(U) M_Y(V)
= \prod\limits_{n=0}^{\infty} \dfrac{\det( 1 - q^n U \otimes V)}{\det( 1 - q^n t U \otimes V)} = \prod\limits_{n=0}^{\infty} \prod\limits_{i,j} \dfrac{1 - q^n U_i V_j}{1 - q^n t U_i V_j}
\end{align}
Here
\begin{align}
g_Y = \prod\limits_{(i,j) \in Y} \frac{1 - t^{Y^T_j-i+1} q^{Y_i-j} }{1 - t^{Y^T_j-i} q^{Y_i-j+1} }
\end{align}
is a normalization factor; note, that it corresponds to the large $N$ limit of $G_Y$.

\paragraph{q,t duality} is a property of Macdonald polynomials. Let ${\hat M}_Y(U)=M_Y(U; t^{-1},q^{-1}).$ Then
\begin{align}
{\hat M}_{Y}(p) \equiv M_{Y}(p)\Big|_{q \leftrightarrow t} = \dfrac{1}{g_{Y_T}} M_{Y^T}\left( p_k = - \dfrac{1-q^k}{1-t^k} p_k \right)
\end{align}
implying that there is no preferred choice of $q,t$. Note, that it can be only formulated if Macdonald polynomials are expressed as functions of the power sums $p_k = {\rm Tr} U^k$.

\paragraph{The inversion.} An important operation defined on the space of symmetric polynomials is the inversion
\begin{align}
i f(p) = f \big( -p \big)
\end{align}
Note that, just as in the previous case, the operation involves a transformation of the variables $p_k$, that cannot be simply expressed in terms of matrix variables $U$.

\paragraph{Dual Cauchy identity} is another version of the Cauchy identity, that features the dual (hatted) Macdonald polynomials:
\begin{align}
\sum\limits_{Y} {\hat M}_{Y^T}(U) M_Y(V)
= \det( 1 + U \otimes V )
\end{align}
It is a consequence of the original Cauchy identity and the $q,t$ duality transformation. Note, that the r.h.s. is independent of $q,t$; hence, also valid for Schur functions.

\paragraph{Product coefficients,} also known as \textbf{Verlinde coefficients} or generalized \textbf{Littlewood - Richardson coefficients}, are the expansion coefficients of a product of two Macdonald polynomials in the same basis:
\begin{align}
M_{Y_1}(U) M_{Y_2}(U) = \sum\limits_{|Y_3| = |Y_1| + |Y_2|} \ N_{Y_1, Y_2}^{Y_3} M_{Y_3}(U)
\end{align}

\paragraph{Skew Macdonald polynomials} $M_{Y/R}$, labeled by two Young diagrams $Y$ and $R$, are a slight generalization of Macdonald polynomials, that arise in many applications, in particular, in construction of the refined topological vertex. They are defined as the following expansion coefficients:
\begin{align}
M_Y(U,V) = \sum\limits_{|R| \leq |Y|} M_{Y/R}(U) M_{R}(V)
\end{align}
Here the "comma" notation means the concatenation of the two sets of eigenvalues, $(U,V) = (U_1,\ldots,U_N,V_1,\ldots,V_M)$. In terms of power sums, this implies $p_k(U,V) = p_k(U) + p_k(V)$. A straightforward calculation shows that skew Macdonald polynomials are given by the following explicit sum, involving the product coefficients:
\begin{align}
M_{Y/R}(U) = \sum\limits_{|A| = |Y|-|R|} N^{Y}_{A,R} \ \dfrac{g_Y}{g_A g_R} \ M_{A}(U)
\end{align}

\paragraph{The shift factor} $v = \sqrt{q/t}$ is a convenient shorthand; this combination becomes 1 in the Schur case $q = t$, and often appears in various contexts that involve generalization from Schur to Macdonald theory.

\section{Analytic continuation of ${\cal O}$-propagators }

\paragraph{Proposition.} Let ${\cal O}_{q\bar q}$ be defined for $V_i > 1$ by an infinite product
$$
{\cal O}_{q\bar q}(t^{\rho} q^{Q}, V) = \prod\limits_{n = 0}^{\infty} \prod\limits_{i=1}^{\infty} \prod\limits_{j} \dfrac{1 - q^n t^{1/2-i} q^{Q_i} V_j^{-1}}{1 - q^n t^{3/2-i} q^{Q_i} V_j^{-1}}
$$
Then, its analytic continuation to $V_i < 1$ is given by ${\cal O}_{q\bar q}(v^2 t^{-\rho} q^{-Q}, V^{-1})$.

\paragraph{Proof.} The problem is to analytically continue
$$
{\cal O}_{q\bar q}(t^{\rho} q^{Q}, V) = \prod\limits_{n = 0}^{\infty} \prod\limits_{i=1}^{\infty} \prod\limits_{j} \dfrac{1 - q^n t^{1/2-i} q^{Q_i} V_j^{-1}}{1 - q^n t^{3/2-i} q^{Q_i} V_j^{-1}}
$$
to $V < 1$, to be able to do a series expansion in positive powers of $V$. We do this in two steps. First, if $Q = \varnothing$, then
$$
{\cal O}_{q\bar q}(t^{\rho}, V) = \prod\limits_{n = 0}^{\infty} \prod\limits_{i=1}^{\infty} \prod\limits_{j} \dfrac{1 - q^n t^{1/2-i} V_j^{-1}}{1 - q^n t^{3/2-i} V_j^{-1}} = \prod\limits_{j} \prod\limits_{n = 0}^{\infty} \dfrac{1}{1 - q^n t^{1/2} V_j} = \prod\limits_{j} \Psi\big( t^{1/2} V_j^{-1} \big)^{-1}
$$
where $\Psi(x) = \prod_n (1 - q^n x)$ is the special function known as quantum dilogarithm. The analytic continuation of this special function from small to large $x$ is well known:
$$
\Psi(x) \ \ \ \mapsto \ \ \ \Psi(q x^{-1})
$$
Consequently,
$$
{\cal O}_{q\bar q}(t^{\rho}, V) \ \ \ \mapsto \ \ \ {\cal O}_{q\bar q}(v^2 t^{-\rho}, V^{-1})
$$
It is now easy to show that ${\cal O}_{q\bar q}(t^{\rho} q^Q, V)$ has the same analytic continuation, as ${\cal O}_{q\bar q}(t^{\rho}, V)$:
\begin{align}
\dfrac{{\cal O}_{q\bar q}(t^{\rho} q^{Q}, V)}{{\cal O}_{q\bar q}(t^{\rho}, V)}
& \ = \ \prod\limits_{n = 0}^{\infty} \prod\limits_{i=1}^{l(Q)} \prod\limits_{j} \dfrac{1 - q^n t^{1/2-i} q^{Q_i} V_j^{-1}}{1 - q^n t^{3/2-i} q^{Q_i} V_j^{-1}} \dfrac{1 - q^n t^{3/2-i} V_j^{-1}}{1 - q^n t^{1/2-i} V_j^{-1}} \\
& \ = \ \prod\limits_{n = 0}^{\infty} \prod\limits_{i=1}^{l(Q)} \prod\limits_{j} \dfrac{1 - q^{-n} t^{-1/2+i} q^{-Q_i} V_j^{+1}}{1 - q^{-n} t^{-3/2+i} q^{-Q_i} V_j^{+1}} \dfrac{1 - q^{-n} t^{-3/2+i} V_j^{+1}}{1 - q^{-n} t^{-1/2+i} V_j^{+1}} \\
& \ = \ \dfrac{{\cal O}_{q\bar q}(v^2 t^{-\rho} q^{-Q}, V^{-1})}{{\cal O}_{q\bar q}(v^2 t^{-\rho}, V^{-1})}
\end{align}
Hence, eq. (\ref{ac})
$$
{\cal O}_{q\bar q}(t^{\rho} q^Q, V) \ \ \ \mapsto \ \ \ {\cal O}_{q\bar q}(v^2 t^{-\rho} q^Q, V^{-1})
$$
is valid. This completes the proof.

\paragraph{Proposition.} Propagators ${\cal O}_{q t}$ and ${\cal O}_{q {\bar q}}$ are related by
\begin{align}
{\cal O}_{q {\bar q}}( t^{\rho} q^{Q} , V) = {\cal O}_{q t}( q^{-\rho} t^{-Q^T}, V)^{-1}
\label{OOIdentity}
\end{align}

\paragraph{Proof.} We have
$$
{\cal O}_{q t}( q^{-\rho} t^{-Q^T}, V)^{-1} = \prod\limits_{i = 1}^{\infty} \prod\limits_{j} \dfrac{1}{1 - v^{-1} q^{-1/2 + i} q^{-Q^T_i} V_j^{-1}} = \exp\left( \sum\limits_{k = 1}^{\infty} \dfrac{v^{-k}}{k} p_k\big(q^{-\rho} t^{-Q^T}\big) p_k(V^{-1}) \right)
$$
$$
{\cal O}_{q {\bar q}}( t^{\rho} q^{Q} , V) = \prod\limits_{n = 0}^{\infty} \prod\limits_{i=1}^{\infty} \prod\limits_{j} \dfrac{1 - q^n t^{1/2-i} q^{Q_i} V_j^{-1}}{1 - q^n t^{3/2-i} q^{Q_i} V_j^{-1}} = \exp\left( \sum\limits_{k = 1}^{\infty} \dfrac{-1}{k} \dfrac{1-t^k}{1-q^k} \ p_k\big(t^{\rho} q^{Q}\big) p_k(V^{-1}) \right)
$$
Relation (\ref{OOIdentity}) then follows from the elementary identity
$$
p_k\big(q^{-\rho} t^{-Q^T}\big) = - v^k \dfrac{1-t^k}{1-q^k} p_k\big(t^{\rho} q^{Q}\big)
$$
This completes the proof.

\paragraph{Corollary.} The analytic continuation of ${\cal O}_{q t}(t^{\rho} q^{Q}, V)$ to $V_i < 1$ is given by ${\cal O}_{q t}(v^2 t^{-\rho} q^{-Q}, V^{-1})$.

\section{Derivation of the refined vertex}

As we proved, the analytic continuation of the vertex amplitude has a form
\begin{align}
{\cal C}(V_1, V_2, V_3)=
\sum\limits_{Q_3} M_{Q_3}(t^{\rho})
&{\cal O}_{q\bar q}(v^2 t^{-\rho} q^{-Q_3}, V_1^{-1})
{\cal O}_{qq}(t^{\rho} q^{Q_3}, V_2)\;
\cr& \times {\cal O}_{{q}, { q}}\big( V_1, V_2 \big)\;{M}_{Q_3}(V_3^{-1})/g_{Q_3}
\end{align}
Now we can expand it in Macdonald functions. For this, we use
$$
{\cal O}_{qq}(x, y) =\prod_{i,\alpha}\prod_n {(1-q^n t x_i y^{-1}_j) \over (1-q^n x_i y^{-1}_j) }=\sum_{R} M_R(x)M_R(y^{-1})/g_R
$$
and
$$
{\cal O}_{q{\bar q}}(x, y) = \prod_{i,\alpha}\prod_n {(1-q^n x_i y^{-1}_j) \over (1-q^nt x_i y^{-1}_j) }=\sum_{R} {iM}_R(x)M_R(y^{-1})/g_R
$$
where the operation $i$  is an involution that takes elementary symmetric functions $p_n(x)$ to $-p_n(x)$. As a symmetric function, the Macdonald polynomials have expansion in sums of powers of $p_n$, so the involution on them is defined.
This gives
\begin{align*}
{\cal C}(V_1, V_2, V_3) =
\sum\limits_{A,B,R,Q_3}
M_{Q_3}(t^{\rho}) {M}_{Q_3}(V_3^{-1})/g_{Q_3} \
{v^{2|A|}} iM_{A}(t^{-\rho} q^{-Q_3}) M_A(V_1)/{g_{A}} \ \times
\end{align*}
\begin{align}
\ \qquad\qquad\qquad\times  M_B(t^{\rho} q^{Q_3}) M_B(V_2^{-1})/{g_{B}} \  M_R(V_1) M_R(V_2^{-1})/{g_{R}} \
\end{align}
To expand the topological vertex from this sum, we simply expand the products of Macdonald polynomials into linear combination thereof, using the formulas
$$
M_A(V_1) M_R(V_1) = \sum\limits_{Q_1} N_{A,R}^{Q_1} M_{Q_1}(V_1)
$$
$$
M_B(V_2^{-1}) M_R(V_2^{-1}) = \sum\limits_{Q_2} N_{B,R}^{Q_2} M_{Q_2}(V_2^{-1})
$$
where $N$ are the multiplication constants of Macdonald polynomials. This gives
\begin{align*}
{\cal C}(V_1, V_2, V_3) =
\sum\limits_{Q_1,Q_2,Q_3}
{v^{2|Q_1|}}\
{\cal C}_{Q_1,Q_2,Q_3} \
iM_{Q_1}(V_1)/{g_{Q_1}} M_{Q_2}(V_2^{-1})/{g_{Q_2}} {M}_{Q_3}(V_3^{-1})/{g_{Q_3}}
\end{align*}
with coefficients given by
\begin{align}
{\cal C}_{Q_1,Q_2,Q_3} =
\sum\limits_{A,B,R} \
\dfrac{v^{-2|R|}}{g_{R}} \ \dfrac{g_{Q_1} g_{Q_2}}{g_{A} g_{B}}
\ N_{A,R}^{Q_1} \ N_{B,R}^{Q_2} \
M_{Q_3}(t^{\rho}) \ iM_{A}(t^{-\rho} q^{-Q_3}) \ M_B(t^{\rho} q^{Q_3})
\end{align}
After this is done, it is convenient to remove the intermediate sums over the indices $A$ and $B$, by summing them up with the help of formulas for skew Macdonald polynomials,
$$
iM_{Q/ R} \equiv \sum\limits_{B} \ {\cal N}_{B,R}^{Q} \ \dfrac{g_Q}{g_R g_B} \ iM_B
$$
$$
M_{Q / R} \equiv \sum\limits_{A} \ {\cal N}_{A,R}^{Q} \ \dfrac{g_Q}{g_R g_A} \ M_A
$$
This gives
\begin{align}
{\cal C}_{Q_1,Q_2,Q_3} =
\sum\limits_{R} \ v^{-2|R|} \ g_R \ iM_{A/R}(t^{-\rho} q^{-Q_3}) \ M_{B/R}(t^{\rho} q^{Q_3}) \ M_{Q_3}(t^{\rho})
\end{align}
that is our final result.
\section{Comparison to the Iqbal-Kozcaz-Vafa vertex}

In this section, we compare our vertex
$$
{\cal C}_{Q_1,Q_2,Q_3} =
\sum\limits_{R} \ v^{-2|R|} \ g_R \ iM_{Q_1/R}(t^{-\rho} q^{-Q_3}) \ M_{Q_2/R}(t^{\rho} q^{Q_3}) \ M_{Q_3}(t^{\rho})
$$
to the Iqbal-Kozcaz-Vafa vertex
$$
{\cal C}^{IKV}_{Q_1,Q_2,Q_3} = \sum\limits_{R} \ v^{-|R|} \ s_{Q_1^T/R}(t^{-\rho} q^{-Q_3}) \ s_{Q_2/R}(q^{-\rho} t^{-Q_3^T}) M_{Q_3}(t^{\rho})
$$
As explained in section 4, these two vertices are simply the expansion coefficients of one and the same vertex amplitude
\begin{align}
{\cal C}^{\prime}(V_1, V_2, V_3) =
\sum\limits_{Q_3}{ 1\over g_{Q_3}} M_{Q_3}(t^{\rho})
& {\cal O}_{q t}(v^2 t^{-\rho} q^{-Q_3}, V_1^{-1})
{\cal O}_{q {\bar q}}(t^{\rho} q^{Q_3}, V_2)\;
\cr& \times {\cal O}_{q, t}\big( V_1, V_2 \big)\;{M}_{Q_3}(V_3^{-1})
\end{align}
in two different bases.

\paragraph{Proposition I.} Expanding the vertex amplitude ${\cal C}^{\prime}(V_1, V_2, V_3)$ in the basis of Macdonald functions, we get our vertex:
$$
{\cal C}'(V_1, V_2, V_3)=\sum_{Q_1,Q_2,Q_3} v^{|Q_1|} \ {\cal C}_{Q_1,Q_2,Q_3} \ i{\hat M}_{Q_1^T}(V_1) iM_{Q_2}(V_2^{-1})/g_{Q_2} M_{Q_3}(V_3^{-1})/g_{Q_3}
$$

\paragraph{Proof.} The proof is a straightforward exercise: using the expansions
$$
{\cal O}_{q t}(v^2 t^{-\rho} q^{-Q_3}, V_1^{-1}) =
\sum\limits_{A} v^{|A|}
iM_{A}(t^{-\rho} q^{-Q_3}) {\hat iM}_{A^T}(V_1)
$$
$$
{\cal O}_{q, t}\big( V_1, V_2 \big) =
\sum\limits_{R} v^{-|R|}
{\hat iM}_{R^T}(V_1) iM_{R}(V_2^{-1})
$$
$$
{\cal O}_{q {\bar q}}(t^{\rho} q^{Q_3}, V_2) =
\sum\limits_{B}
\dfrac{1}{g_B} M_{B}(t^{\rho} q^{Q_3}) iM_{B}(V_2^{-1})
$$
multiplying the Macdonald polynomials
\begin{align*}
{\hat iM}_{A^T}(V_1) {\hat iM}_{Y^T}(V_1) & \ = \ \sum\limits_{Q_1} \ {\hat N}^{Q_1^T}_{A^T,R^T} \ {\hat iM}_{Q_1^T}(V_1) = \\
& \ = \  \sum\limits_{Q_1} \ \dfrac{g_{Q_1}}{g_{A} g_{R}} \ N^{Q_1}_{A,R} \ {\hat iM}_{Q_1^T}(V_1)
\end{align*}
$$
iM_{B}(V_2^{-1}) iM_{R}(V_2^{-1}) = \sum\limits_{Q_2} \ N_{B,R}^{Q_2} \ iM_{Q_2}(V_2^{-1})
$$
and converting the sums over auxillary partitions $A,B$ to skew Macdonald polynomials
$$
\sum\limits_{A} \ N_{A,R}^{Q_1} \ \dfrac{v^{|A|}}{g_A} iM_{A}(t^{-\rho} q^{-Q_3}) = v^{|Q_1|-|R|} \dfrac{g_R}{g_{Q_1}} \ iM_{Q_1/R}(t^{-\rho} q^{-Q_3})
$$
$$
\sum\limits_{B} \ N_{B,R}^{Q_2} \ \dfrac{1}{g_B} M_{B}(t^{\rho} q^{Q_3}) = \dfrac{g_R}{g_{Q_2}} \ M_{Q_2/R}(t^{\rho} q^{Q_3})
$$
we finally find
$$
{\cal C}'(V_1, V_2, V_3)=\sum_{Q_1,Q_2,Q_3} \dfrac{v^{|Q_1|}}{g_{Q_3} g_{Q_2}} \ {\cal C}_{Q_1,Q_2,Q_3}(q,t) \ i{\hat M}_{Q_1^T}(V_1) iM_{Q_2}(V_2^{-1}) M_{Q_3}(V_3^{-1}).
$$
where ${\cal C}_{Q_1,Q_2,Q_3}$ is our vertex.

\paragraph{Proposition II.} Expanding the vertex amplitude ${\cal C}^{\prime}(V_1, V_2, V_3)$ in the basis of Schur functions, we get the IKV vertex:
$$
{\cal C}'(V_1, V_2, V_3)=\sum_{Q_1,Q_2,Q_3} \ v^{|Q_1|-|Q_2|} \ {\cal C}^{IKV}_{Q_1,Q_2,Q_3} \ s_{Q_1}(V_1) s_{Q_2}(V_2^{-1}) M_{Q_3}(V_3^{-1})
$$

\paragraph{Proof.} Completely similar. Using the expansions
$$
{\cal O}_{q t}( v^2 t^{-\rho} q^{-Q_3}, V_1^{-1}) =
\sum\limits_{A} v^{|A|}
s_{A}(t^{-\rho} q^{-Q_3}) s_{A^T}(V_1)
$$
$$
{\cal O}_{q, t}\big( V_1, V_2 \big) =
\sum\limits_{R} v^{-|R|}
s_{R^T}(V_1) s_{R}(V_2^{-1})
$$
$$
{\cal O}_{q {\bar q}}(t^{\rho} q^{Q_3}, V_2) =
{\cal O}_{q t}( q^{-\rho} t^{-Q_3^T}, V_2)^{-1} =
\sum\limits_{B} v^{-|B|}
s_{B}(q^{-\rho} t^{-Q_3^T}) s_{B}(V_2^{-1})
$$
multiplying the Schur polynomials
\begin{align*}
s_{A^T}(V_1) s_{R^T}(V_1) & \ = \ \sum\limits_{Q_1} \ {\widetilde N}^{Q_1}_{A^T,R^T} \ s_{Q_1}(V_1) = \\
& \ = \  \sum\limits_{Q_1} \ {\widetilde N}^{Q_1^T}_{A,R} \ s_{Q_1}(V_1)
\end{align*}
$$
s_{B}(V_2^{-1}) s_{R}(V_2^{-1}) = \sum\limits_{Q_2} \ {\widetilde N}_{B,R}^{Q_2} \ s_{Q_2}(V_2^{-1})
$$
and converting the sums over auxillary partitions $A,B$ to skew Schur polynomials
$$
\sum\limits_{A} \ v^{|A|} \ {\widetilde N}_{A,R}^{Q_1^T} \ s_{A}(t^{-\rho} q^{-Q_3}) = v^{|Q_1|-|R|} \ s_{Q_1^T/R}(t^{-\rho} q^{-Q_3})
$$
$$
\sum\limits_{B} \ v^{-|B|} \ N_{B,R}^{Q_2} \ s_{B}(q^{-\rho} t^{-Q_3^T}) = v^{-|Q_2|+|R|} \ s_{Q_2/R}(q^{-\rho} t^{-Q_3^T})
$$
we finally find
$$
{\cal C}'(V_1, V_2, V_3) = \sum_{Q_1,Q_2,Q_3} \ v^{|Q_1|-|Q_2|} \ {\cal C}^{IKV}_{Q_1,Q_2,Q_3} \ s_{Q_1}(V_1) s_{Q_2}(V_2^{-1}) s_{Q_3}(V_3^{-1}).
$$
where ${\cal C}^{IKV}_{Q_1,Q_2,Q_3}$ is the Iqbal-Kozcaz-Vafa vertex.


\end{document}